\documentclass[12pt]{article}
\usepackage{a4wide}
\usepackage{latexsym}
\usepackage{amsmath}
\usepackage{amsfonts}
\usepackage{amscd}
\usepackage{amsthm}
\usepackage{amssymb}
\usepackage{cite}
\usepackage{hyperref}
\usepackage{placeins}
\usepackage{mathdots}
\usepackage{axodraw2}
\usepackage{tcolorbox}
\tcbuselibrary{breakable}
\usepackage{nomencl}

\usepackage{color}

\usepackage{pslatex}
\usepackage{graphicx}
\usepackage[latin1,utf8]{inputenc}
\usepackage[T1]{fontenc}

\allowdisplaybreaks

\newcommand{\bq}{\begin{eqnarray}}
\newcommand{\eq}{\end{eqnarray}}
\newcommand{\eps}{\varepsilon}

\newcommand{\NB}{N_B}

\newcommand{\NF}{N_F}

\newcommand{\NV}{n}

\newcommand{\ND}{N_D}

\newcommand{\differentialform}{\Psi}

\newcommand{\Hgen}{H}
\newcommand{\Agen}{\Omega}

\newcommand{\Fcomb}{F_{\mathrm{comb}}}
\newcommand{\Fgeom}{F_{\mathrm{geom}}}

\newcommand{\preabs}{C_{\mathrm{abs}}}
\newcommand{\prerel}{C_{\mathrm{rel}}}
\newcommand{\preclutch}{C_{\mathrm{clutch}}}
\newcommand{\prebaikov}{C_{\mathrm{Baikov}}}

\newcommand{\Divisor}{P}

\newcommand{\absmu}{|\mu|}

\newcommand{\persymmetricmatrix}{S}

\newcommand{\myhat}[1]{#1}

\newcounter{algocounter}

\theoremstyle{plain}

\makenomenclature

\begin{document}

\thispagestyle{empty}

\begin{flushright}
\end{flushright}

\vspace{1.5cm}

\begin{center}
  {\Large\bf 
 Intersection matrices associated to geometric-ordered bases of Feynman integrals
 \\
  }
  \vspace{1cm}
  {\large Iris Bree${}^{a}$,
          Federico Gasparotto${}^{b}$,
          Sebastian~P\"ogel${}^{c}$, 
          Xing~Wang${}^{d}$,
          Stefan~Weinzierl${}^{a}$ and
          Xiaofeng~Xu${}^{e}$
\\
  \vspace{1cm}
      {\small \em ${}^{a}$ PRISMA Cluster of Excellence, Institut f{\"u}r Physik, Staudinger Weg 7,} \\
      {\small \em Johannes Gutenberg-Universit{\"a}t Mainz, D-55099 Mainz, Germany}\\
  \vspace{2mm}
      {\small \em ${}^{b}$ Bethe Center for Theoretical Physics, Universität Bonn, D-53115 Bonn, Germany} \\
  \vspace{2mm}
      {\small \em ${}^{c}$ Department of Astrophysics, University of Zurich, Winterthurerstrasse 190, 8057 Zurich, Switzerland} \\
  \vspace{2mm}
      {\small \em ${}^{d}$ School of Science and Engineering,} \\
      {\small \em The Chinese University of Hong Kong, Shenzhen, 518172 Guangdong, China} \\
  \vspace{2mm}
      {\small \em ${}^{e}$ Department of Physics, Xiamen University, Xiamen, 361005, China} 
  } 
\end{center}

\vspace{2cm}

\begin{abstract}\noindent
  {
In integration-by-parts reduction of Feynman integrals, the order relation in the Laporta algorithm
determines a set of master integrals.
In this paper we investigate the intersection matrices of the integrands of the master integrals that 
are obtained from a geometric order relation.
With an appropriate definition of integrands and their duals, we find that the intersection matrices 
are simpler than expected: 
For a filtration-compatible basis, the entries of the intersection matrix are Laurent polynomials 
in the dimensional regularisation parameter $\varepsilon$.
For an $\varepsilon$-factorised basis, the entries are instead integers,
up to an overall power of $\varepsilon$,
if the boundary values for the auxiliary functions of the rotation are chosen appropriately. This has practical consequences: We can systematically eliminate certain auxiliary transcendental functions, introduced in going from a filtration-compatible basis to an $\varepsilon$-factorised basis. By exploiting some structural properties of the intersection matrices, we provide an algorithm that performs this elimination while minimising the number of required calculations.
   }
\end{abstract}

\vspace*{\fill}

\newpage

\section{Introduction}
\label{sect:intro}

Precision calculations in high-energy physics rely on our ability to compute Feynman integrals.
To advance these abilities we need to keep the size of intermediate expressions under control.
We may obtain more efficient algorithms through a better understanding of the mathematics underlying Feynman integrals.
In particular, the close connection between integration-by-parts identities 
and twisted cohomology \cite{Mastrolia:2018uzb,Frellesvig:2019kgj,Frellesvig:2019uqt,Mizera:2019vvs,Weinzierl:2020xyy,Frellesvig:2020qot,Caron-Huot:2021xqj,Caron-Huot:2021iev,Chestnov:2022alh,Chestnov:2022xsy,Fontana:2023amt,Brunello:2023rpq,Brunello:2024tqf,Chen:2020uyk,Chen:2022lzr,Jiang:2023qnl,Jiang:2023oyq,Huang:2026xnq}
plays a crucial role.
Recently, we proposed a new order relation for the Laporta algorithm \cite{e-collaboration:2025frv,Bree:2025tug}.
This order relation orders Feynman integrals within a sector by properties of algebraic geometry
(essentially the number of residues and the number of poles).
In this paper we investigate the intersection matrices of the integrands of the master integrals within 
a given sector.
The entries of these matrices are intersection numbers, i.e. pairings between an integrand and a dual integrand.
Given a basis of integrands, denoted by $\langle \differentialform_i |$ in the following, we first have to
define a basis of the dual twisted cohomology group.
The standard dual basis $| \Lambda_j \rangle$ is defined by the property that the intersection matrix is the unit matrix:
\bq
 \langle \differentialform_i | \Lambda_j \rangle & = & \delta_{ij}.
\eq
In this case, there is not much more to be said about the intersection matrix (it is by definition the unit matrix), 
except that all complications
are buried in the computation of the standard dual basis $| \Lambda_j \rangle$.

We are interested in a basis of the dual twisted cohomology group that can easily be derived
from a basis of the twisted cohomology group.
Specifically, this dual basis $| \differentialform_j^\vee \rangle$ can 
be obtained from $\langle \differentialform_j |$ via the substitution $\eps \rightarrow - \eps$.
We call this basis the $\eps$-dual basis, in order to distinguish it from the standard dual basis 
mentioned above.

There is one complication that needs to be addressed from the beginning:
Feynman integrals may have symmetries of the form
\bq
\label{symmetry}
 I_{\nu_1 \dots \nu_N} & = & I_{\nu_{\sigma(1)} \dots \nu_{\sigma(N)}},
\eq
where the $\nu_i$ denote powers of propagators in the integral, and $\sigma$ is a permutation of $\{1,\dots,N\}$.
While the integrals exhibit such symmetries, the associated integrands typically do not.
In this case, the dimension of the twisted cohomology group 
(the space of integrands modulo integration-by-parts identities) can be larger than the dimension of the 
space spanned by the master integrals (the space of Feynman integrals modulo
integration-by-parts identities and symmetry relations).
The intersection number 
\bq
 \langle \differentialform_i | \differentialform_j^\vee \rangle,
\eq
is invariant under integration-by-parts identities, and therefore well-defined on the cohomology 
classes; it is however in general not invariant under symmetry relations.
If we add to $\differentialform_i$
a term which integrates to zero, the intersection number might change.
For this reason we cannot simply lift the intersection number of two integrands to an ``intersection number'' of two Feynman integrals.
This is not well-defined,
as it depends which symmetry-relation equivalent integrand we pick.
One resolution is trivialising the action of the symmetry relations:
The symmetries of eq.~(\ref{symmetry}) induce an action on the integrands and we may define 
a symmetrised integrand $\differentialform_i^{\mathrm{sym}}$ by averaging over the group action \cite{Gasparotto:2023roh,Duhr:2026elp}.
The intersection numbers of these symmetrised integrands are now well-defined at the level of Feynman integrals,
as the dependence on the initial seed for the symmetrisation drops out.
This allows us to define intersection numbers at the level of Feynman integrals.

With these preparations we investigate the bases returned by a recent algorithm \cite{e-collaboration:2025frv,Bree:2025tug}.
This algorithm consists of two steps. 
After step $1$, one obtains a basis $J$, whose differential equation
is compatible with a filtration and involves only rational functions.
In step $2$, we construct a rotation to a basis $K$, whose differential equation is in $\eps$-factorised form
and may involve transcendental functions.
These transcendental objects are introduced via auxiliary functions in the rotation, and are 
determined by differential equations.
In solving these differential equations there is a freedom in choosing 
boundary values. For an $\eps$-factorised differential equation any choice that leads
to an invertible rotation between the bases $J$ and $K$ is allowed. 
The auxiliary transcendental functions are inevitable to achieve an $\eps$-factorised differential equation. They also
appear in other approaches \cite{Gorges:2023zgv,Duhr:2025lbz,Yang:2025ofz,Chen:2025hzq,Dlapa:2022wdu}
and are also called $\eps$-functions \cite{Duhr:2025kkq} or $G$-functions \cite{Gorges:2023zgv,Duhr:2025lbz}\footnote{By contrast, the construction of~\cite{Chaubey:2025adn} avoids the presence of such functions, at the cost of giving up $\eps$-factorisation.}.
A discussion of the auxiliary functions in terms of leading singularities can be found in \cite{Forner:2026vby}.
 
We study the intersection matrices for the bases $J$ and $K$ on the maximal cut.
The restriction to the maximal cut avoids relative twisted cohomology and allows us to work entirely with
(non-relative) twisted cohomology. 
For the basis $J$ we expect the intersection matrix to be rational in the kinematic variables $x$
and the dimensional regularisation parameter $\eps$.
We observe that the intersection matrix is in fact much simpler than that: 
The entries are Laurent polynomials in $\eps$.
For the basis $K$ we expect 
the intersection matrix to be constant with respect to the kinematic variables $x$ \cite{Duhr:2024xsy,Duhr:2025xyy},
if an appropriate choice of the boundary values for the auxiliary functions has been made.
Again, we find a stronger result: The entire $\eps$-dependence is given by a prefactor $\eps^{-\NV}$
and the remaining entries are integers.
The non-trivial part of this statement is the proportionality to a power of $\eps$. Given that the entries of the intersection matrix are rational,
the integer condition follows easily from an appropriate rescaling.
Note that the statement above depends on an appropriate choice of the boundary values for the auxiliary functions.
For a generic choice, the differential equation for the basis $K$ will be $\eps$-factorised, but the intersection matrix
will not necessarily be constant.
We may turn this argument around and choose the integration constants for the auxiliary functions
such that the intersection matrix is constant.
The requirement of constant intersection numbers is a (global) property that we may require in addition
to an $\eps$-factorised form for the differential equation.
We remark that requiring at most simple poles in the differential equation is a property that in general only holds
locally, but not globally \cite{Frellesvig:2023iwr}.

There is one practical application: 
Step $2$ of the algorithm of refs.~\cite{e-collaboration:2025frv,Bree:2025tug} 
introduces auxiliary functions, whose number we would like to minimise.
It is known that self-duality \cite{Pogel:2024sdi,Duhr:2024xsy,Duhr:2024rxe} or constant intersection numbers \cite{Duhr:2025xyy}
can be used to obtain algebraic relations between these functions,
which in turn can be used to eliminate a subset.
This approach has been applied in the past for 
concrete examples \cite{Pogel:2022yat,Pogel:2022ken,Pogel:2022vat,Duhr:2024uid,Duhr:2025kkq}.
The results of our paper are two-fold:
First, we show that the requirement of constant intersection numbers provides more constraints than the requirement of self-duality. 
Second, building upon earlier proposals of using the constancy of intersection numbers \cite{Duhr:2024xsy,Duhr:2024rxe,Duhr:2025xyy},
we present a systematic algorithm that eliminates algebraically redundant auxiliary functions on the maximal cut.
Relations among the auxiliary transcendental functions will in general involve certain rational functions.
The method we present computes these rational functions entirely within the field of rational functions (i.e. independently of any transcendental auxiliary function).
This is done by first computing the intersection numbers associated to the (rational) basis $J$.
The computation of these intersection numbers is achieved through the solution of suitable differential equations \cite{Matsubara-Heo:2019} (see also \cite{Chestnov:2022alh,Duhr:2025kkq}).
\begin{figure}
\begin{center}
\includegraphics[scale=0.9]{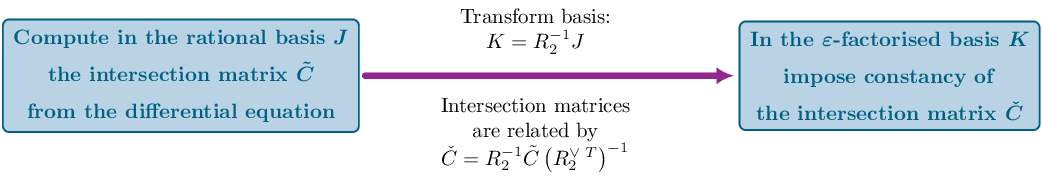}
\end{center}
\caption{
The flow chart for the algorithm to eliminate algebraically redundant auxiliary functions.
}
\label{fig:flow_chart}
\end{figure}
We show how the properties of the basis $J$, as well as certain symmetries of the intersection matrices, facilitate this step in practice. 
The central idea of the algorithm is illustrated in fig.~\ref{fig:flow_chart}.

This paper is organised as follows:
In the next section we introduce the setup.
In section~\ref{sect:symmetry_intersection_matrix} we discuss symmetries of the intersection matrix.
In section~\ref{sect:compatibility} we discuss the relation between
$\differentialform_i$ and $\differentialform_i^{\mathrm{sym}}$.
In section~\ref{sect:algorithm} we provide an algorithm, which eliminates algebraically redundant auxiliary functions on the maximal cut.
In section~\ref{sect:examples} we present several examples.
Finally, our conclusions are given in section~\ref{sect:conclusions}.
In appendix~\ref{sect:map_iota} we provide technical details on the map to compact support.
In appendix~\ref{sect:invariance} we provide additional details on the invariance of intersection numbers under certain transformations.


\printnomenclature

\nomenclature[s000]{For the convenience of the reader, we list the main symbols used in this article:}{}
\nomenclature[s010]{\bf Differential forms:}{}
\nomenclature[s011]{$\differentialform$}{A generic differential form.}
\nomenclature[s012]{$\tilde{\differentialform}$}{A differential form representing a basis element of the rational basis $J$.}
\nomenclature[s013]{$\check{\differentialform}$}{A differential form representing a basis element of the $\eps$-factorised basis $K$.}
\nomenclature[s014]{$\hat{\differentialform}$}{A differential form representing a basis element of a basis satisfying strong self-duality.}

\nomenclature[s020]{\bf Connection matrices:}{}
\nomenclature[s021]{$A$}{A generic connection matrix.}
\nomenclature[s022]{$\tilde{A}$}{The connection matrix of the rational basis $J$.}
\nomenclature[s023]{$\check{A}$}{The connection matrix of the $\eps$-factorised basis $K$.}
\nomenclature[s024]{$\hat{A}$}{The connection matrix of a basis satisfying strong self-duality.}

\nomenclature[s030]{\bf Intersection matrices:}{}
\nomenclature[s031]{$C$}{A generic intersection matrix.}
\nomenclature[s032]{$\tilde{C}$}{The intersection matrix of the rational basis $J$.}
\nomenclature[s033]{$\check{C}$}{The intersection matrix of the $\eps$-factorised basis $K$.}
\nomenclature[s034]{$\hat{C}$}{The intersection matrix of a basis satisfying strong self-duality.}

\nomenclature[s040]{\bf Rescaled intersection matrices:}{}
\nomenclature[s041]{$\overline{C}$}{The rescaled version of $C$, defined by  $\overline{C} = \eps^{\NV} C$.}
\nomenclature[s042]{$\overline{\tilde{C}}$}{The rescaled version of $\tilde{C}$, defined by  $\overline{\tilde{C}} = \eps^{\NV} \tilde{C}$.}
\nomenclature[s043]{$\overline{\check{C}}$}{The rescaled version of $\check{C}$, defined by  $\overline{\check{C}} = \eps^{\NV} \check{C}$.}
\nomenclature[s044]{$\overline{\hat{C}}$}{The rescaled version of $\hat{C}$, defined by  $\overline{\hat{C}} = \eps^{\NV} \hat{C}$.}

\section{The setup}
\label{sect:setup}

\subsection{General definitions}
\label{sect:general_def}

We consider a fibre bundle with fibre ${\mathbb C}{\mathbb P}^{\NV}$ and base space ${\mathbb C}{\mathbb P}^{\NB}$.
We denote the homogeneous coordinates on the fibre by $[z_0:z_1:\dots:z_{\NV}]$
and the homogeneous coordinates on the base by $[x_0:x_1:\dots:x_{\NB}]$.
The differential in the fibre is denoted by
\bq
\label{def_d_F}
 d_F & = & \sum\limits_{j=0}^{\NV} dz_j \frac{\partial}{\partial z_j}
 + \sum\limits_{j=0}^{\NV} d\overline{z}_j \frac{\partial}{\partial \overline{z}_j},
\eq
the differential on the base space is denoted by
\bq
\label{def_d_B}
 d_B & = & \sum\limits_{j=0}^{\NB} dx_j \frac{\partial}{\partial x_j}
 + \sum\limits_{j=0}^{\NB} d\overline{x}_j \frac{\partial}{\partial \overline{x}_j}.
\eq
For holomorphic functions, the derivatives with respect to the anti-holomorphic variables can be neglected.
This case will be of main interest. 
The anti-holomorphic derivatives will only be relevant for the map $\iota_\omega$,
which maps a differential form to an equivalent differential form with compact support. 
This map appears only in the definition of intersection numbers.
We denote the dimensional regularisation parameter by $\eps \in {\mathbb C}$.
We consider a set of $(1+\ND)$ homogeneous rational functions $\Divisor_0\left(z,x\right), \Divisor_1\left(z,x\right), \dots, \Divisor_{\ND}\left(z,x\right)$,
which are polynomials in the variables $z$.
We denote by $d_j$ the degree of the polynomial $\Divisor_j\left(z,x\right)$ with respect to the variables $z$.
With respect to the variables $x$, the functions $\Divisor_j\left(z,x\right)$ are homogeneous of degree $0$.
The functions $\Divisor_j\left(z,x\right)$ are independent of $\eps$.
These functions define the twist
\bq
\label{def_twist}
 U & = &
 \prod\limits_{j=0}^{\ND} \Divisor_j^{\alpha_j},
\eq
where the exponents are of the form
\bq
\label{minimal_twist_condition}
 \alpha_j \; = \; 
 \frac{1}{2} \left( a_j + b_j \eps \right)
 & \mbox{with} &
 a_j \; \in \; \{-1,0\}, \;\;\; b_j \; \in \; {\mathbb Z}
\eq
and
\bq
 \frac{1}{2}\sum\limits_{j=0}^{\ND} a_j d_j \; \in \; {\mathbb Z}, & \hspace*{5mm} & \sum\limits_{j=0}^{\ND} b_j d_j \; = \; 0.
\eq
We further require $\alpha_j \notin {\mathbb Z}$, i.e. if $a_j=0$ then $b_j \neq 0$.
The twist function is homogeneous in the variables $z$. We denote by $d_U$ the degree of homogeneity of $U$ with respect
to the variables $z$.
We define $I_{\mathrm{odd}}^0$ as the set of indices for which $a_j$ is odd
and $I_{\mathrm{even}}^0$ as the set of indices for which $a_j$ is even.
The product of all odd polynomials will play an important role and we therefore give it a name:
\bq
 \Divisor_{\mathrm{odd}} & = & \prod\limits_{j \in I_{\mathrm{odd}}^0} \Divisor_j.
\eq
We denote by $d_{\mathrm{odd}}$ the degree of the polynomial $\Divisor_{\mathrm{odd}}$ with respect to the variables $z$.
We have
\bq
 d_U & = & - \frac{1}{2} d_{\mathrm{odd}}.
\eq
At fixed $x$,
each polynomial defines a variety in ${\mathbb C}{\mathbb P}^{\NV}$ and we set
\bq
 D_j \; = \; \{ \; z \in \mathbb{C} \mathbb{P}^{\NV} \; | \; \Divisor_j\left(z\right) = 0 \; \},
\eq
and 
\bq
\label{def_divisior_D}
 D & = & 
 \bigcup\limits_{j=0}^{\ND} D_j.
\eq
On ${\mathbb C}{\mathbb P}^{\NV} - D$ we consider the differential forms
\bq
\label{def_input_data}
 \differentialform_{\mu_0 \dots \mu_{\ND}}\left[Q\right]
 & = &
 C\left(\eps,\mu,a,b\right)
 \;
 U 
 \;
 \myhat{\Phi}_{\mu_0 \dots \mu_{\ND}}\left[Q\right]
 \eta,
\eq
where $C(\eps,\mu,a,b)$ is a prefactor independent of $z$, $\myhat{\Phi}_{\mu_0 \dots \mu_{\ND}}\left[Q\right]$ is given by
\bq
 \myhat{\Phi}_{\mu_0 \dots \mu_{\ND}}\left[Q\right]
 & = &
 \frac{Q}{\prod\limits_{j=0}^{\ND} \Divisor_j^{\mu_j}},
\eq
where $\mu_j \in {\mathbb N}_0$ and $Q=Q(z,x,\eps)$ is a homogeneous polynomial of degree
\bq
\label{def_d_Q}
 d_Q & = &
 \sum\limits_{j=0}^{\ND} \mu_j d_j - d_U - \NV - 1
\eq
in the variables $z$.
$\eta$ is the standard $\NV$-form on ${\mathbb C}{\mathbb P}^{\NV}$ defined by
\bq
\label{def_eta}
 \eta
 & = &
 \sum\limits_{j=0}^{\NV} (-1)^{j} \; z_j \; dz_0 \wedge ... \wedge \widehat{dz_j} \wedge ... \wedge dz_{\NV},
\eq
where the hat indicates that the corresponding term is omitted.
The prefactor\footnote{
In the context of Feynman integrals the prefactor is given as
\bq
 C\left(x,\eps,\mu,a,b\right)
 & = &
 \prebaikov\left(x,\eps\right) \cdot \preabs\left(x,\eps\right) \cdot \prerel\left(\eps,\mu,a,b\right) \cdot \preclutch\left(\eps,\mu\right),
\eq
where $\preabs \cdot \prebaikov$ is pure of transcendental weight zero.
This product is not relevant for the discussion here and we factor it out.
As any rational number is pure of transcendental weight zero, we actually must factor it out in the context of this paper in order
to have an integer condition on the intersection numbers.
}
is 
given as a product 
\bq
 C\left(\eps,\mu,a,b\right)
 & = &
 \prerel\left(\eps,\mu,a,b\right) \cdot \preclutch\left(\eps,\mu\right).
\eq
The relative prefactor $\prerel$ is given by
\bq
\label{def_pre_rel}
 \prerel\left(\eps,\mu,a,b\right)
 & = &
 \prod\limits_{i \in I_{\mathrm{all}}^0} 
 \left( \frac{1}{2} \left( a_i + b_i \eps \right)\right)_{\mu_i},
\eq
with $(a)_n=\Gamma(a+1)/\Gamma(a+1-n)$ being the falling factorial.
The clutch prefactor $\preclutch$ is given by
\bq
 \preclutch\left(\eps,\mu\right)
 & = &
 \eps^{-\absmu},
 \;\;\;\;\;\;\;\;\;
 \left| \mu \right|
 \; = \;
 \sum\limits_{i \in I_{\mathrm{all}}^0} \mu_i.
\eq
We denote the set of all such differential forms $\differentialform_{\mu_0 \dots \mu_{\ND}}\left[Q\right]$
by $\Agen^{\NV}$. This is a vector space of countable dimension.

The $n$-forms $\differentialform_{\mu_0 \dots \mu_{\ND}}\left[Q\right] \in \Agen^{\NV}$ are holomorphic on ${\mathbb C}{\mathbb P}^{\NV} - D$
and therefore closed on ${\mathbb C}{\mathbb P}^{\NV} - D$.
Modding out the exact ones defines the twisted cohomology group $H^\NV$,
whose elements we denote by $\langle \differentialform |$:
\bq
 \left\langle \differentialform \right|
 & \in &
 \Hgen^{\NV}.
\eq
The differential form $\differentialform$ is a representative for the cohomology class $\langle \differentialform |$.
Any two representatives differ by an exact form.
We may further consider the dual space, which we denote by
\bq
 \left| \differentialform^\vee \right\rangle
 & \in &
 \left( \Hgen^{\NV} \right)^\vee.
\eq
There is additionally the twisted homology group, which we denote by
\bq
 \left| {\mathcal C} \right\rangle
 & \in &
 \Hgen_{\NV}.
\eq
Finally, there is the dual of the twisted homology group, which we denote by
\bq
 \left\langle {\mathcal C}^\vee \right|
 & \in &
 \left(\Hgen_{\NV}\right)^\vee.
\eq
The dimensions of these four spaces are equal and finite:
\bq
 \dim \Hgen^{\NV} \; = \; \dim \left( \Hgen^{\NV} \right)^\vee \; = \; \dim \Hgen_{\NV} \; = \; \dim \left(\Hgen_{\NV}\right)^\vee \; = \; \NF.
\eq
There are pairings between the bra- and the ket-objects, which define for the bases of these four spaces four $(\NF \times \NF)$-matrices:
\bq
\begin{array}{lllll}
 P_{ij} & = & \left\langle \differentialform_i | {\mathcal C}_j \right\rangle & : & \mbox{period matrix}, \\
 P^{\vee T}_{ij} & = & \left\langle {\mathcal C}_i^\vee | \differentialform_j^\vee \right\rangle & : & \mbox{dual period matrix}, \\
 H_{ij} & = & \left\langle {\mathcal C}_i^\vee | {\mathcal C}_j \right\rangle & : & \mbox{intersection matrix of cycles}, \\
 C_{ij} & = & \left\langle \differentialform_i | \differentialform_j^\vee \right\rangle & : & \mbox{intersection matrix of cocycles}. \\
\end{array}
\eq
All matrices have full rank and are therefore invertible.
Riemann's twisted bilinear relations~\cite{cho1995} provide a relation among these four pairings:
\bq
\label{eq:RTPR}
 P H^{-1} P^{\vee T} & = & \left( 2 \pi i \right)^{\NV} C.
\eq

\subsection{Bases and dual bases}
\label{sect:bases}

From a basis of $\Hgen^{\NV}$ we may obtain a basis of $( \Hgen^{\NV} )^\vee$ by the substitution $\eps \rightarrow -\eps$.
This is the most natural choice in the context of this paper.
We therefore define for any quantity $X(\eps)$
\bq
\label{substitution_eps}
 X^\vee\left(\eps\right) & = & X\left(-\eps\right).
\eq
Let us discuss this in detail.
We recall from eq.~(\ref{def_twist}) that the twist function is given by
\bq
 U & = &
 \prod\limits_{j=0}^{\ND} \Divisor_j^{\frac{1}{2} \left( a_j + b_j \eps \right)},
\eq
with $a_j \in \{-1,0\}$.
Obviously,
\bq
 U^{-1} & = &
 \prod\limits_{j=0}^{\ND} \Divisor_j^{-\frac{1}{2} \left( a_j + b_j \eps \right)}.
\eq
Then $U^\vee$ is given by
\bq
 U^\vee & = &
 \prod\limits_{j=0}^{\ND} \Divisor_j^{\frac{1}{2} \left( a_j - b_j \eps \right)},
\eq
obtained from $U$ by the substitution $\eps \rightarrow - \eps$, see eq.~(\ref{substitution_eps}).
Clearly,
\bq
\label{relation_inverse_twist}
 U^{-1} & = & \Divisor_{\mathrm{odd}} U^\vee.
\eq
As $U^{-1}$ and $U^\vee$ differ only by a rational function, 
whose denominator involves at most only polynomials $\Divisor_j$, 
it follows that $( \Hgen^{\NV} )^\vee$ 
can be studied either with the twist function $U^{-1}$ or with
the twist function $U^\vee$.
We may compute the dimension of $( \Hgen^{\NV} )^\vee$ and a basis of $( \Hgen^{\NV} )^\vee$
by using either twist function. 
The reason is the following: 
The exact forms are $d_F \Xi$, where $\Xi$ contains the twist function and a rational function.
The assignment of a factor $\Divisor_{\mathrm{odd}}$ either to the twist function or to the rational function 
will not affect the exact forms and therefore will not affect the cohomology.
If we use $U^\vee$ as twist function, we may recycle for the computation of $( \Hgen^{\NV} )^\vee$
the results for the computation of $\Hgen^{\NV}$ through
the substitution $\eps \rightarrow -\eps$ and a transposition.
Only in the intersection matrix $C$ we have to take an extra factor $\Divisor_{\mathrm{odd}}$ into account.

We now make a specific choice for a basis of $H^\NV$: 
We consider for $H^\NV$ an $\eps$-factorised 
basis $\langle \check{\differentialform}_i |$ (with $1 \le i \le \NF$)
constructed with the help of the algorithms of refs.~\cite{e-collaboration:2025frv,Bree:2025tug}.
This basis satisfies
\bq
\label{eps_factorised_diff_eq}
 d_B \left\langle \check{\differentialform}_i \right|
 & = &
 \eps \sum\limits_{j=1}^{\NF} \check{A}_{ij}\left(x\right) \left\langle \check{\differentialform}_j \right|.
\eq
With the standard assumption that the boundaries of the cycles are within the divisor $D$, it follows that the period matrix satisfies
\bq
 d_B \check{P} & = & \eps \check{A} \check{P}.
\eq
We denote by $| \check{\differentialform}_j^\vee \rangle$ the basis of $( \Hgen^{\NV} )^\vee$ obtained from
$\langle \check{\differentialform}_j |$ by the substitution $\eps \rightarrow - \eps$.
More explicitly, if
\bq
 \check{\differentialform}_j
 & = &
 C_j\left(\eps,\mu,a,b\right)
 \;
 U 
 \;
 \frac{Q_j\left(z,x,\eps\right)}{\prod\limits_{k=0}^{\ND} \Divisor_k^{\mu_{jk}}}
 \eta,
\eq
then
\bq
 \check{\differentialform}_j^\vee
 & = &
 C_j\left(-\eps,\mu,a,b\right)
 \;
 U^\vee
 \;
 \frac{Q_j\left(z,x,-\eps\right)}{\prod\limits_{k=0}^{\ND} \Divisor_k^{\mu_{jk}}}
 \eta
 \; = \;
 C_j\left(-\eps,\mu,a,b\right)
 \;
 U^{-1}
 \;
 \frac{Q_j\left(z,x,-\eps\right)}{\Divisor_{\mathrm{odd}} \prod\limits_{k=0}^{\ND} \Divisor_k^{\mu_{jk}}}
 \eta.
\eq
Note that an additional factor $\Divisor_{\mathrm{odd}}$ appears in the denominator, if we write $\check{\differentialform}_j^\vee$
with the twist function $U^{-1}$.
This follows from eq.~(\ref{relation_inverse_twist}).
Note further that the additional factor has no influence on the prefactor $C_j\left(-\eps,\mu,a,b\right)$.
This basis satisfies
\bq
 d_B \left| \check{\differentialform}_j^\vee \right\rangle
 & = & 
 - \eps \sum\limits_{i=1}^{\NF} \left| \check{\differentialform}_i^\vee \right\rangle \check{A}_{ji}\left(x\right). 
\eq
If the boundaries of the dual cycles are also contained within the divisor $D$, it follows that the dual period matrix satisfies
\bq
 d_B \check{P}^{\vee T} & = & - \eps \check{P}^{\vee T} \check{A}^T.
\eq
It follows from eq.~(\ref{eq:RTPR}) (and from the fact that $d_B H =0$) that
\bq
\label{diff_intersection_matrix}
 d_B \check{C} & = & \eps \left( \check{A} \check{C} - \check{C} \check{A}^T \right).
\eq
This equation is a special case of an equation called the secondary equation in ref.~\cite{Matsubara-Heo:2019}.

Besides the $\eps$-factorised basis, we consider a second basis.
The algorithm of refs.~\cite{e-collaboration:2025frv,Bree:2025tug} constructs in step 1 a
basis $\langle \tilde{\differentialform}_i |$ of $\Hgen^{\NV}$, 
which we denote with a tilde and whose differential
equation is in Laurent polynomial form:
\bq
 d_B \left\langle \tilde{\differentialform}_i \right|
 \; = \;
 \sum\limits_{j=1}^{\NF} \tilde{A}_{ij}\left(\eps,x\right) \left\langle \tilde{\differentialform}_j \right|,
 & &
 \tilde{A}_{ij}\left(\eps,x\right)
 \; = \;
 \sum\limits_{k=-(\absmu_i-\absmu_j)}^1
 \eps^k \tilde{A}^{(k)}_{ij}\left(x\right).
\eq
The entries of $\tilde{A}^{(k)}_{ij}\left(x\right)$ are rational differential one-forms.
The basis $\langle \tilde{\differentialform}_i |$ does not involve any auxiliary functions.
After step $2$ in the algorithm of refs.~\cite{e-collaboration:2025frv,Bree:2025tug}
one obtains a basis $\langle \check{\differentialform}_i |$ that satisfies the $\eps$-factorised differential equation of eq.~(\ref{eps_factorised_diff_eq}).
The basis $\langle \check{\differentialform}_i |$ may involve auxiliary functions.
The two bases are related by a rotation $R_2$
\bq
 \left\langle \tilde{\differentialform}_i \right|
 & = &
 \sum\limits_{j=1}^{\NF} 
 \left(R_2\right)_{ij} \left\langle \check{\differentialform}_j \right|.
\eq
Associated to the basis $\langle \tilde{\differentialform}_j |$ of $\Hgen^{\NV}$ is the basis
$| \tilde{\differentialform}_j^\vee \rangle$ of $(\Hgen^{\NV})^\vee$, again obtained by the substitution $\eps \rightarrow -\eps$.
This basis satisfies the differential equation
\bq
 d_B \left| \tilde{\differentialform}_j^\vee \right\rangle
 & = & 
 \sum\limits_{i=1}^{\NF} \left| \tilde{\differentialform}_i^\vee \right\rangle \tilde{A}_{ji}\left(-\eps,x\right). 
\eq
We denote the corresponding intersection matrix by $\tilde{C}$, e.g.
\bq
 \tilde{C}_{ij}
 & = &
 \left\langle \tilde{\differentialform}_i | \tilde{\differentialform}_j^\vee \right\rangle.
\eq
This intersection matrix satisfies the differential equation
\bq
\label{diff_eq_tildeC}
 d_B \tilde{C} & = & \tilde{A} \tilde{C} + \tilde{C} \tilde{A}^{\vee T}.
\eq
This equation is called the secondary equation in ref.~\cite{Matsubara-Heo:2019}.
In this reference it is also shown that this equation has under certain assumptions up to constant multiplication a unique rational solution (theorem~(2.1) in ref.~\cite{Matsubara-Heo:2019}).
In order to keep the discussion simple, we will assume in the following that the assumptions 
of this theorem are met\footnote{We are aware of Feynman integrals, where these assumptions are not met. In these cases the master integrals on the maximal cut split into smaller blocks and we may apply the theorem to the smaller blocks on the diagonal.}.

With this theorem and as $\tilde{A}(\eps,x)$ is known, we may use the differential equation~(\ref{diff_eq_tildeC}) to determine the intersection matrix $\tilde{C}$
up to an overall prefactor $f(\eps)$ that is independent of $x$, 
but possibly dependent on $\eps$.
In order to determine $f(\eps)$ it is sufficient to determine one non-zero entry of $\tilde{C}$ by other means.

For later purposes, we define rescaled versions of the intersection matrices $C$, $\check{C}$ and $\tilde{C}$:
We set
\bq
 \overline{C} \; = \; \eps^{\NV} C,
 \;\;\;\;\;\;
 \overline{\check{C}} \; = \; \eps^{\NV} \check{C},
 \;\;\;\;\;\;
 \overline{\tilde{C}} \; = \; \eps^{\NV} \tilde{C}.
\eq
A priori, the matrices $\overline{\check{C}}$ and $\overline{\tilde{C}}$ may still depend on $\eps$.
We will later see in concrete examples that $\overline{\check{C}}$ is independent of $\eps$ 
(and that the entries of $\overline{\tilde{C}}$ or $\tilde{C}$ are Laurent polynomials in $\eps$).
We introduce these rescaled intersection matrices as they have slightly simpler symmetry properties.

\subsection{Self-duality}

In this subsection we start a small detour and review self-duality, discussed in refs.~\cite{Pogel:2024sdi,Duhr:2024xsy}.
We will later show that as far as the elimination of the auxiliary functions is concerned, self-duality is superseded by the constancy of intersection numbers.
For this reason self-duality will not appear in the algorithm for the elimination of the auxiliary functions.

Let $\persymmetricmatrix$ denote the $(\NF \times \NF)$-matrix
\bq
 \persymmetricmatrix & = & 
 \left(\begin{array}{ccccc}
 0 & \dots & \dots & 0 & 1 \\
 \vdots & & \iddots & 1 & 0 \\
 \vdots & \iddots & \iddots & \iddots & \vdots \\
 0 & 1 & \iddots & & \vdots \\
 1 & 0 & \dots & \dots & 0 \\
 \end{array}\right).
\eq
The strong version of self-duality \cite{Pogel:2024sdi} states that there exists a basis 
$\langle \hat{\differentialform}_i |$ of $H^\NV$ with differential equation
\bq
 d_B \left\langle \hat{\differentialform}_i \right|
 & = &
 \eps \sum\limits_{j=1}^{\NF} \hat{A}_{ij}\left(x\right) \left\langle \hat{\differentialform}_j \right|,
\eq
such that
\bq
\label{strong_self_duality}
 \hat{A} & = & \persymmetricmatrix \hat{A}^T \persymmetricmatrix^{-1}.
\eq
The weak version of self-duality \cite{Duhr:2024xsy} states that for the basis $\langle \check{\differentialform}_i |$ of $H^\NV$ 
there exists a constant symmetric matrix $\hat{C}$ such that
\bq
\label{weak_self_duality}
 \check{A} & = & \hat{C} \check{A}^T \hat{C}^{-1}.
\eq
If the intersection matrix $\check{C}$ is constant (i.e. $d_B \check{C}=0$), we may take $\hat{C}=\check{C}$ and 
eq.~(\ref{weak_self_duality}) follows directly from eq.~(\ref{diff_intersection_matrix}).
Note that weak self-duality requires in addition that $\check{C}$ is symmetric.
In section~\ref{sect:symmetry_intersection_matrix} we show that if the $\eps$-dependence of
the intersection matrix is simply given by a prefactor $\eps^{-\NV}$, then $\check{C}$ 
is necessarily symmetric.

If we relate the bases $\langle \hat{\differentialform}_i |$ and $\langle \check{\differentialform}_j |$
by a constant rotation
\bq
 \left\langle \hat{\differentialform}_i \right| 
 & = & \sum\limits_{j=1}^{\NF} R_{ij} \left\langle \check{\differentialform}_j \right|
\eq
and $R$ is chosen such that
\bq
 R \check{C} R^T & = & \persymmetricmatrix
\eq
it is easily shown that the weak version of self-duality implies the strong version~\cite{Duhr:2024xsy}.
It is usually the case that the entries of $\check{C}$ are rational numbers, while the entries of the rotation matrix $R$
may involve algebraic extensions.

\section{Symmetry of the intersection matrix}
\label{sect:symmetry_intersection_matrix}

Let us study the symmetries of the intersection matrix $C$ for the case that 
the basis $| \differentialform_j^\vee \rangle$ of $( \Hgen^{\NV} )^\vee$
is obtained from the basis $\langle \differentialform_i |$ of $\Hgen^{\NV}$
by $\eps \rightarrow -\eps$ (and division by $P_{\mathrm{odd}}$, if it refers to the twist function $U^{-1}$ instead of $U^\vee$). 

We first investigate how intersection numbers transform if we change the twist by a specific rational function.
The specific rational functions we would like to consider are of the form
\bq
\label{def_T_shift}
 T & = &
 \prod\limits_{j=0}^{\ND} \Divisor_j^{\beta_j},
 \;\;\;\;\;\;
 \beta_j \; \in \; {\mathbb Z}.
\eq
These functions have the property that in the denominator of $T$ and $T^{-1}$ only polynomials from the twist function occur.
The differential form $\differentialform$ is invariant under the simultaneous change
\bq
\label{shift_twist}
 U \rightarrow U T,
 & &
 \myhat{\Phi} \rightarrow T^{-1} \myhat{\Phi}.
\eq
Similarly, the dual differential form $\differentialform^\vee$ is invariant under the simultaneous change
\bq
\label{shift_twist_dual}
 U^\vee \rightarrow U^\vee T^{-1},
 & &
 \myhat{\Phi}^\vee \rightarrow T \myhat{\Phi}^\vee.
\eq
We set
\bq
 \omega
 \; = \;
 d_F \ln U,
 \;\;\;
 \omega^\vee
 \; = \;
 d_F \ln U^\vee,
 \;\;\;
 \kappa
 \; = \;
 d_F \ln T,
 \;\;\;
 \kappa_{\mathrm{odd}}
 \; = \;
 d_F \ln \Divisor_{\mathrm{odd}}.
\eq
From eq.~(\ref{relation_inverse_twist}) we have
\bq
 - \omega & = & \omega^\vee + \kappa_{\mathrm{odd}}.
\eq
The intersection numbers are defined by \cite{cho1995,Aomoto:book}
\bq
\label{def_intersection_number}
 C_{ij}
 \;= \;
 \left\langle \differentialform_i \right. \left| \differentialform_j^\vee \right\rangle
 & = &
 \frac{C_iC_j^\vee}{\left(2\pi i\right)^n}
 \int
  \iota_\omega\left(\myhat{\Phi}_i \eta\right)
  \wedge
  \iota_{-\omega}\left(\frac{\myhat{\Phi}_j^\vee \eta}{\Divisor_{\mathrm{odd}}}\right)
 \nonumber \\
 & = &
 \frac{C_iC_j^\vee}{\left(2\pi i\right)^n}
 \int
  \iota_\omega\left(\frac{Q_i \eta}{\prod\limits_{k=0}^{\ND} \Divisor_k^{\mu_{ik}}} \right)
  \wedge
  \iota_{-\omega}\left(\frac{Q_j^\vee \eta}{\Divisor_{\mathrm{odd}} \prod\limits_{k=0}^{\ND} \Divisor_k^{\mu_{jk}}} \right),
\eq
where
\begin{align}
 C_i & = C\left(\eps,\mu_i,a,b\right),
 & 
 C_j^\vee & = C\left(-\eps,\mu_j,a,b\right),
 \nonumber \\
 Q_i & = Q_i\left(z,x,\eps\right),
 &
 Q_j^\vee & = Q_j\left(z,x,-\eps\right).
\end{align}
The map $\iota_\omega$ sends its argument to its compactly supported version,
and similarly for $\iota_{-\omega}$.
In the literature one often finds formulae where either $\iota_\omega$ is applied to the first factor
or $\iota_{-\omega}$ to the second factor.
One can show that applying the map to both factors yields the same result.
The symmetric version is slightly advantageous for our purposes.
In appendix~\ref{sect:map_iota} we provide additional information on the map $\iota_{\omega}$.

In the following we will assume that the intersection numbers are invariant
under the simultaneous transformations of eq.~(\ref{shift_twist}) and eq.~(\ref{shift_twist_dual}):
\bq
\label{rescaling_equation}
 \frac{1}{\left(2\pi i\right)^n}
 \int \iota_\omega\left(\myhat{\Phi}_i \eta\right) \wedge \iota_{-\omega}\left(\frac{\myhat{\Phi}_j^\vee \eta}{\Divisor_{\mathrm{odd}}}\right)
 & = &
 \frac{1}{\left(2\pi i\right)^n}
 \int \iota_{\omega+\kappa}\left(\frac{\myhat{\Phi}_i \eta}{T}\right) \wedge \iota_{-\omega-\kappa}\left(\frac{T \myhat{\Phi}_j^\vee \eta}{\Divisor_{\mathrm{odd}}}\right).
\eq
We provide more details on this formula in appendix~\ref{sect:invariance}.
We may specialise to $\kappa=\kappa_{\mathrm{odd}}$:
\bq
 \frac{1}{\left(2\pi i\right)^n}
 \int \iota_\omega\left(\myhat{\Phi}_i \eta\right) \wedge \iota_{-\omega}\left(\frac{\myhat{\Phi}_j^\vee \eta}{\Divisor_{\mathrm{odd}}}\right)
 & = &
 \frac{1}{\left(2\pi i\right)^n}
 \int \iota_{-\omega^\vee}\left(\frac{\myhat{\Phi}_i \eta}{\Divisor_{\mathrm{odd}}}\right) \wedge \iota_{\omega^\vee}\left(\myhat{\Phi}_j^\vee \eta\right).
\eq
Let us now compare $C_{ij}$ and $C_{ji}$. We have
\bq
 C_{ij}
 & = &
 \frac{C_i C_j^\vee}{\left(2\pi i\right)^n}
 \int 
 \iota_\omega\left(\myhat{\Phi}_i \eta\right)
 \wedge 
 \iota_{-\omega}\left(\frac{\myhat{\Phi}_j^\vee \eta}{\Divisor_{\mathrm{odd}}}\right),
 \nonumber \\
 C_{ji}
 & = &
 \frac{C_j C_i^\vee}{\left(2\pi i\right)^n}
 \int 
 \iota_\omega\left(\myhat{\Phi}_j \eta\right)
 \wedge 
 \iota_{-\omega}\left(\frac{\myhat{\Phi}_i^\vee \eta}{\Divisor_{\mathrm{odd}}}\right)
 \; = \;
 \frac{\left(-1\right)^{\NV} C_i^\vee C_j }{\left(2\pi i\right)^n}
 \int 
 \iota_{\omega^\vee}\left(\myhat{\Phi}_i^\vee \eta\right)
 \wedge 
 \iota_{-\omega^\vee}\left( \frac{\myhat{\Phi}_j \eta}{\Divisor_{\mathrm{odd}}} \right).
 \nonumber
\eq
From these expressions we 
find that
\bq
 C_{ji} & = & \left(-1\right)^{\NV} C_{ij}^\vee.
\eq
In terms of the rescaled intersection matrix $\overline{C} = \eps^{\NV} C$ we have
\bq
\label{symmetry_rescaled_intersection_matrix}
 \overline{C} & = & \overline{C}^{\vee T}.
\eq
Let us decompose $\overline{C}$ into a part $\overline{E}$, which is even under $\eps \rightarrow - \eps$, and a part $\overline{O}$, which is odd:
\bq
 \overline{C} \; = \; \overline{E} + \overline{O},
 \;\;\;\;\;\;
 \overline{E} \; = \; \frac{1}{2}\left( \overline{C} + \overline{C}^\vee \right),
 \;\;\;\;\;\;
 \overline{O} \; = \; \frac{1}{2}\left( \overline{C} - \overline{C}^\vee \right).
\eq
Then eq.~(\ref{symmetry_rescaled_intersection_matrix}) implies that
\bq
 \overline{E} \; = \; \overline{E}^T,
 & &
 \overline{O} \; = \; -\overline{O}^T,
\eq
i.e. $\overline{E}$ is symmetric with respect to transposition, while $\overline{O}$ is anti-symmetric.
In particular, if $\overline{\check{C}}$ is independent of $\eps$, it must be even under $\eps \rightarrow - \eps$
and therefore symmetric.
Therefore, under the assumption that $\overline{\check{C}}$ is independent of $\eps$,
the possibility of an anti-symmetric $\overline{\check{C}}$ discussed in ref.~\cite{Duhr:2024xsy} is excluded.

\section{Compatibility of intersection numbers with symmetry relations from integration}
\label{sect:compatibility}

In the application towards Feynman integrals we are primarily concerned with the integrals, 
not integrands. 
As $\dim \Hgen^{\NV} = \dim \Hgen_{\NV} = \NF$ we have $\NF$ independent integration cycles ${\mathcal C}_j$ on the maximal cut.
We are primarily interested in one specific integration cycle ${\mathcal C}_{\mathrm{Feynman}}$, 
the one which derives from the loop momentum representation of Feynman integrals, where the integration is over all real loop momenta with deformations towards
complex loop momenta according to Feynman's $i \delta$-prescription.
The cycle ${\mathcal C}_{\mathrm{Feynman}}$ can in principle be expressed as a linear combination of the independent cycles ${\mathcal C}_j$.
The exact form of this linear combination is not relevant in the following argument.
The essential point is that we are only interested in one specific integration cycle.

Through this section we therefore denote a Feynman integral as
\bq
 I & = & \left\langle \differentialform | {\mathcal C}_{\mathrm{Feynman}} \right\rangle,
\eq
with one specific cycle ${\mathcal C}_{\mathrm{Feynman}}$.
It may happen that for a basis element $\langle \differentialform_i |$ of $\Hgen^{\NV}$ we have   
\bq
\label{symmetry_relation}
 \left\langle \differentialform_i | {\mathcal C}_{\mathrm{Feynman}} \right\rangle & = & 0,
\eq
although $\langle \differentialform_i | \neq 0$, as it is a basis element of $\Hgen^{\NV}$.
Eq.~(\ref{symmetry_relation}) is called a symmetry relation.
In order to study this issue in more detail, we denote the vector space of Feynman integrals 
on the maximal cut modulo linear relations by $V^{\NV}$
and the twisted cohomology group related to the integrands in a Baikov representation by $\Hgen^{\NV}$.
Both are finite-dimensional vector spaces.
There is an injective linear map
\bq
\label{def_iota}
 \iota & : & V^{\NV} \rightarrowtail \Hgen^{\NV}.
\eq
The map $\iota$, 
when applied to a Feynman integral, gives the integrand in the chosen 
Baikov representation\footnote{The map $\iota$ in eq.~(\ref{def_iota}) has nothing to do with the maps $\iota_{\omega}$ and $\iota_{-\omega}$ appearing in eq.~(\ref{def_intersection_number}).}.
In general, the map $\iota$ will not be surjective, as there can be symmetry relations and super-sectors, as discussed in refs.~\cite{e-collaboration:2025frv,Bree:2025tug}.
We split the basis of $\Hgen^{\NV}$ into three sets:
Elements of the first set vanish after integration against $| {\mathcal C}_{\mathrm{Feynman}} \rangle$ and correspond to symmetry relations,
elements of the second and third set do not vanish after integration against $| {\mathcal C}_{\mathrm{Feynman}} \rangle$.
The elements of the second set correspond to the Feynman integrals from the sector under consideration, while
elements of the third set correspond to Feynman integrals from super-sectors.
With a slight abuse of notation we denote the vector of elements from the first set by $\langle \differentialform_1 |$,
the vector of elements from the second set by $\langle \differentialform_2 |$,
and the vector of elements from the third set by $\langle \differentialform_3 |$.
As before we denote by $| \differentialform_1^\vee \rangle$, $| \differentialform_2^\vee \rangle$ 
and $| \differentialform_3^\vee \rangle$ the corresponding elements
of $(\Hgen^{\NV})^\vee$, obtained by the substitution $\eps \rightarrow -\eps$ and transposition.
We have the differential equation
\bq
 d_B \left( \begin{array}{c}
  \left\langle \differentialform_1 \right| \\
  \left\langle \differentialform_2 \right| \\
  \left\langle \differentialform_3 \right| \\
 \end{array} \right)
 & = &
 \left( \begin{array}{ccc}
  A_{11} & 0 & 0 \\
  A_{21} & A_{22} & 0 \\
  A_{31} & A_{32} & A_{33} \\
 \end{array} \right)
 \left( \begin{array}{c}
  \left\langle \differentialform_1 \right| \\
  \left\langle \differentialform_2 \right| \\
  \left\langle \differentialform_3 \right| \\
 \end{array} \right).
\eq
The block entry $A_{23}$ is zero, because the third set is a super-sector of the second set.
The block entries $A_{12}$ and $A_{13}$ must be zero for the following reason:
After integration with $| {\mathcal C}_{\mathrm{Feynman}} \rangle$ we obtain
\bq
\label{diff_eq_three_blocks}
 d_B \left( \begin{array}{c}
  0 \\
  \left\langle \differentialform_2 | {\mathcal C}_{\mathrm{Feynman}} \right\rangle \\
  \left\langle \differentialform_3 | {\mathcal C}_{\mathrm{Feynman}} \right\rangle \\
 \end{array} \right)
 & = &
 \left( \begin{array}{ccc}
  A_{11} & 0 & 0 \\
  A_{21} & A_{22} & 0 \\
  A_{31} & A_{32} & A_{33} \\
 \end{array} \right)
 \left( \begin{array}{c}
  0 \\
  \left\langle \differentialform_2 | {\mathcal C}_{\mathrm{Feynman}} \right\rangle \\
  \left\langle \differentialform_3 | {\mathcal C}_{\mathrm{Feynman}} \right\rangle \\
 \end{array} \right).
\eq
As $\langle \differentialform_2 | {\mathcal C}_{\mathrm{Feynman}} \rangle$ and $\langle \differentialform_3 | {\mathcal C}_{\mathrm{Feynman}} \rangle$ 
constitute a basis of the sector and the super-sectors, respectively, they are linearly independent.
Hence it follows that we must have
\bq
 A_{12} \; = \; 0,
 & &
 A_{13} \; = \; 0,
\eq
as otherwise we would have the linear relation
\bq
 A_{12} \left\langle \differentialform_2 | {\mathcal C}_{\mathrm{Feynman}} \right\rangle
 +
 A_{13} \left\langle \differentialform_3 | {\mathcal C}_{\mathrm{Feynman}} \right\rangle 
 & = & 0.
\eq
We denote by 
\bq
 I & = & \left\langle \differentialform_2 | {\mathcal C}_{\mathrm{Feynman}} \right\rangle
\eq
the vector of master integrals in the sector under consideration. From eq.~(\ref{diff_eq_three_blocks}) we have
\bq
 d_B I & = & A_{22} I.
\eq
We may compute an intersection matrix $C_{\mathrm{Feynman}}$ (up to an overall prefactor $f(\eps)$, which is not relevant in the discussion here)
by solving the differential equation
\bq
\label{diff_eq_C_Feynman}
 d_B C_{\mathrm{Feynman}} & =& A_{22} C_{\mathrm{Feynman}} + C_{\mathrm{Feynman}} A_{22}^{\vee T}.
\eq
We may then ask the question whether the matrix $C_{\mathrm{Feynman}}$ is the intersection matrix
\bq
 \left\langle \differentialform_{2, i} | \differentialform_{2, j}^\vee \right\rangle.
\eq
The subtle and maybe surprising answer is no. In order to see this, we compute the full intersection matrix
\bq
 C & = &
 \left( \begin{array}{ccc}
  C_{11} & C_{12} & C_{13} \\
  C_{21} & C_{22} & C_{23} \\
  C_{31} & C_{32} & C_{33} \\
 \end{array} \right)
\eq
from
\bq
 d_B C & = & A C + C A^{\vee T}.
\eq
The differential equation for $C_{22}$ reads
\bq
 d_B C_{22} & = & A_{22} C_{22} + C_{22} A_{22}^{\vee T} + A_{21} C_{12} + C_{21} A_{21}^{\vee T}.
\eq
This differs from eq.~(\ref{diff_eq_C_Feynman}) if $A_{21} C_{12} + C_{21} A_{21}^{\vee T} \neq 0$.
We may then ask which intersection numbers eq.~(\ref{diff_eq_C_Feynman}) actually computes?
To answer this question we first note that the symmetries of the Feynman integral family form a group $G$.
We assume that $g \in G$ acts linearly on the variables $z$ as $z' = g \cdot z$ and that $\det g = \pm 1$. 
While the Feynman integral is invariant under the action of $g \in G$, the integrand of a Feynman integral is not necessarily invariant.
However, we may form symmetrised integrands by averaging over the group action \cite{Gasparotto:2023roh,Duhr:2026elp}:
\bq
 \left\langle \differentialform_j^{\mathrm{sym}} \right|
 & = & \frac{1}{\left| G \right|} \sum\limits_{g \in G} g \cdot \left\langle \differentialform_j \right|,
 \;\;\;\;\;\;
 j \; \in \; \{2,3\},
\eq
where $g$ acts on $\differentialform_j(z) = f_j(z) d^\NV z$ as
\bq
 g \cdot \differentialform_j(z) & = & 
 \det g \cdot \differentialform_j\left(g^{-1} \cdot z\right)
 \; = \; 
 f_j\left(g^{-1} \cdot z\right) \; d^\NV z.
\eq
The differential equation for the symmetrised differential forms reads
\bq
 d_B \left( \begin{array}{c}
  \left\langle \differentialform_2^{\mathrm{sym}} \right| \\
  \left\langle \differentialform_3^{\mathrm{sym}} \right| \\
 \end{array} \right)
 & = &
 \left( \begin{array}{cc}
  A_{22} & 0 \\
  A_{32} & A_{33} \\
 \end{array} \right)
 \left( \begin{array}{c}
  \left\langle \differentialform_2^{\mathrm{sym}} \right| \\
  \left\langle \differentialform_3^{\mathrm{sym}} \right| \\
 \end{array} \right),
\eq
and it follows that $C_{\mathrm{Feynman}}$ is the intersection matrix of the symmetrised differential forms 
$\langle \differentialform_2^{\mathrm{sym}} |$.

There is an immediate corollary:
Let $I$ be a basis of master integrals for the maximal cut of some sector.
We may consider for the master integrals various integral representations:
the loop momentum representation, the Feynman parameter representation, the Lee-Pomeransky representation, 
the Baikov representation, etc.
In any integral representation, the symmetrised differential forms of
the integrands 
$\left\langle \differentialform^{\mathrm{sym}} \right|$  and $\left| \differentialform^{\mathrm{sym}\; \vee} \right\rangle$ 
will always have (up to an overall normalisation) the intersection matrix $C_{\mathrm{Feynman}}$.

Proof: The differential equation for the intersection matrix $C$ is
\bq
 d_B C_{\mathrm{Feynman}} & =& A C_{\mathrm{Feynman}} + C_{\mathrm{Feynman}} A^{\vee T}.
\eq
The differential equation only depends on $A$, which is independent of the integral representation.
This proves the corollary.

From now on we will always consider the symmetrised integrands 
$\langle \differentialform_i^{\mathrm{sym}} |$ and $| \differentialform_j^{\mathrm{sym} \;\vee}\rangle$
and we will write for simplicity $C$ instead of $C_{\mathrm{Feynman}}$.
In all examples we checked we found that the rescaled intersection matrix
$\overline{\tilde{C}}$ (i.e. the rescaled intersection matrix for the filtration-compatible basis $J$ after step $1$) is a Laurent polynomial in $\eps$
\bq \label{Laurent_C}
 \overline{\tilde{C}}\left(\eps,x\right)
 & = &
 \sum\limits_{k=k_{\mathrm{min}}}^0 \overline{\tilde{C}}^{(k)}\left(x\right) \eps^k,
\eq
with $k_{\mathrm{min}} \ge - \NV$
and that the rescaled intersection matrix $\overline{\check{C}}$ (i.e. the rescaled intersection matrix for the $\eps$-factorised basis $K$ after step $2$)
is independent of $\eps$ and $x$ for
an appropriate choice of the boundary values for the auxiliary functions:
\bq
 \overline{\check{C}}\left(\eps,x\right)
 & = &
 \overline{\check{C}}^{(0)}.
\eq
In section~\ref{sect:symmetry_intersection_matrix} we showed that 
$\overline{\tilde{C}}^{(k)}\left(x\right)$ is symmetric if $k$ is even and anti-symmetric if $k$ is odd.
The matrix $\overline{\check{C}}^{(0)}$ is always symmetric.

\section{The algorithm for eliminating auxiliary functions}
\label{sect:algorithm}

In this section we present an algorithm for the elimination of some of the auxiliary functions introduced in step 2 of the algorithm of
refs.~\cite{e-collaboration:2025frv,Bree:2025tug} on the maximal cut.
This is of particular importance, if the sector has a large number of master integrals.
In this case, the number of auxiliary functions on the maximal cut will be large as well, and we would like to reduce this number.
We may obtain relations among the auxiliary functions by requiring that the intersection matrix $\check{C}$ of the $\eps$-factorised basis $K$ is constant.
The algorithm will verify that the constancy of the intersection numbers can be imposed.
Relations among the auxiliary transcendental function will in general involve certain rational functions.
The method we present computes these rational functions entirely within the field of rational functions (i.e. independently of any transcendental auxiliary function).

We denote by $J$ a basis of master integrals on the maximal cut of a sector of interest after step $1$ of the algorithm of refs.~\cite{e-collaboration:2025frv,Bree:2025tug}.
This basis satisfies a differential equation
\bq
 d_B J & = & \tilde{A}\left(\eps,x\right) J,
\eq
where $\tilde{A}\left(\eps,x\right)$ is compatible with a filtration.
In particular, $\tilde{A}\left(\eps,x\right)$ is in Laurent polynomial form.
The intersection matrix $\tilde{C}$ of this basis can be computed by finding a rational solution to the differential equation~(\ref{diff_eq_tildeC})
\bq
 d_B \tilde{C} & = & \tilde{A} \tilde{C} + \tilde{C} \tilde{A}^{\vee T}.
\eq
We further denote by $K$ the corresponding basis after step $2$ of the algorithm of refs.~\cite{e-collaboration:2025frv,Bree:2025tug}.
The basis $K$ satisfies an $\eps$-factorised  differential equation
\bq
 d_B K & = & \eps \check{A}\left(x\right) K.
\eq
The two bases are related by
\bq
 J & = & R_2 K.
\eq
The intersection matrix $\check{C}$ of this basis is given by
\bq
 \check{C} & = & R_2^{-1} \tilde{C} \left(R_2^{\vee \; T}\right)^{-1}.
\eq
The rotation matrix $R_2$ may involve auxiliary functions.
Step $2$ of the algorithm of refs.~\cite{e-collaboration:2025frv,Bree:2025tug} provides differential equations for all auxiliary functions.
In order to eliminate some of the auxiliary functions, we proceed as follows:
\begin{tcolorbox}[breakable]
\refstepcounter{algocounter}
\label{algo:construct_masters}
{\bf Algorithm \thealgocounter: Elimination of auxiliary functions}

\begin{description}

\item{\bf Input:}
The matrix $\tilde{A}$, the matrix $R_2$ in terms of the auxiliary functions and the differential equations for the auxiliary functions.

\item{\bf Output:}
Algebraic relations among the auxiliary functions, which can be used to eliminate some of the auxiliary functions.

\item{\bf Implementation:}
\begin{enumerate}
\item Compute a matrix $\overline{\tilde{C}}$ up to an overall prefactor $f(\eps)$ by solving the differential equation
\bq
\label{diff_eq_overlineCtilde}
 d_B \overline{\tilde{C}} & = & \tilde{A} \overline{\tilde{C}} + \overline{\tilde{C}} \tilde{A}^{\vee T}.
\eq
Partially fix the prefactor by requiring that $\det \overline{\tilde{C}}$ is non-zero and independent of $\eps$. 
Verify that $\overline{\tilde{C}}$ is rational and a Laurent polynomial in $\eps$ with exponents $\le 0$. If not, throw an exception.
\item Define
\bq
 \overline{\check{C}} & = & R_2^{-1} \overline{\tilde{C}} \left(R_2^{\vee \; T}\right)^{-1}.
\eq
\item Verification step: 
Write
\bq
 \overline{\check{C}} & = & \sum\limits_{k=k_{\mathrm{min}}}^0 \overline{\check{C}}^{(k)} \eps^k.
\eq
For $k$ from $k_{\mathrm{min}}$ to $0$ use 
the differential equations for the auxiliary functions 
and $\overline{\check{C}}^{(k')}_{ij}=0$ for $k'<k$
to verify that
\bq
 d_B \overline{\check{C}}^{(k)} & = & 0.
\eq
If this is not true, throw an exception.
If it is true, we have verified that $\overline{\check{C}}^{(k)}$ is constant.
If $k<0$ set $\overline{\check{C}}^{(k)}_{ij}=0$.
\item 
Choose numbers $N_{ij}$ subject to $\det N \neq 0$ and $N_{ij}=N_{ji}$.
Return the set of algebraic equations
\bq
 \begin{array}{ll}
 \overline{\check{C}}^{(k)}_{ij} = 0, & k<0, \\
 \overline{\check{C}}^{(0)}_{ij} = N_{ij}, & k=0. \\
 \end{array}
\eq
\end{enumerate}
\end{description}
\end{tcolorbox}
A few comments are in order: 
\begin{enumerate}
\item The only non-trivial computation is the integration of the differential equation
in eq.~(\ref{diff_eq_overlineCtilde}).
As the result is expected to be rational, this is usually feasible.
The methods of refs.~\cite{BARKATOU1999547,10.1145/2442829.2442840} are helpful for this task. As a supplement to this, assuming eq.~(\ref{Laurent_C}), the symmetry properties discussed in section~\ref{sect:symmetry_intersection_matrix} facilitate the integration, reducing the number of unknowns. 

\item It is easily shown that the $\eps$-dependence and the $x$-dependence of $\det \overline{\tilde{C}}$
factorise:
From Jacobi's formula for determinants we have
\bq
 d_B \ln \det \overline{\tilde{C}}
 & = &
 \mathrm{Tr}\left( \overline{\tilde{C}}^{-1} d_B \overline{\tilde{C}} \right) 
 \; = \;
 \mathrm{Tr}\left(\tilde{A} + \tilde{A}^{\vee T}\right).
\eq
We assumed that $\tilde{A}$ is compatible with a filtration, which implies that all entries on the diagonal are linear 
in $\eps$. The $\eps^{1}$-terms cancel out on the diagonal in the combination $\tilde{A} + \tilde{A}^{\vee T}$.
Therefore,
$\mathrm{Tr}\left(\tilde{A} + \tilde{A}^{\vee T}\right)$ is independent of $\eps$.
Hence 
\bq
 \det \overline{\tilde{C}} & = & f(\eps) \exp\left( \int \mathrm{Tr}\left(\tilde{A} + \tilde{A}^{\vee T}\right) \right).
\eq

\item In the verification step we need to work bottom-up, i.e. starting with $k=k_{\mathrm{min}}$ and ending with $k=0$.
The information that $\overline{\check{C}}^{(k')}_{ij}=0$ for $k'<k$ is in general required to show
$d_B \overline{\check{C}}^{(k)} = 0$.

\item We will show in explicit examples that the requirement of constant intersection numbers may give more algebraic
relations than the requirement of self-duality.
In the case where all equations are independent and come from the leading order in $1/\eps$, this can be understood as follows:
As $\overline{\check{C}}$ is symmetric, the requirement of constant intersection numbers gives $\NF(\NF+1)/2$ algebraic equations.
On the other hand, self-duality will only give $\NF(\NF-1)/2$ algebraic equations, as
the left-hand side of 
\bq
 \check{A} \overline{\check{C}} - \overline{\check{C}} \check{A}^T & = & 0
\eq
is anti-symmetric.

\item In the case, where the transformation $R_2$ is algebraic (and therefore does not involve auxiliary transcendental functions),
self-duality implies that not all entries of $\check{A}$ are independent.
As an example we consider $\NF=2$. From self-duality we get one algebraic equation:
\bq
\label{relation_Nf_eq_2}
 \overline{\check{C}}_{12} \left( \check{A}_{11} - \check{A}_{22} \right) + \overline{\check{C}}_{22} \check{A}_{12} - \overline{\check{C}}_{11} \check{A}_{21} & = & 0.
\eq
Thus, if self-duality holds, the four entries of $\check{A}$ cannot be independent, but there exist three numbers
\bq
 \overline{\check{C}}_{11}, \overline{\check{C}}_{12}, \overline{\check{C}}_{22}
\eq
such that eq.~(\ref{relation_Nf_eq_2}) holds.

\end{enumerate}

\section{Examples}
\label{sect:examples}

In this section we present a few examples.
The first two examples are rather simple and do not involve auxiliary functions.
They mainly serve to illustrate eq.~(\ref{relation_Nf_eq_2}).
Starting from example $3$ we discuss examples which involve auxiliary functions.
Example $3$ involves an elliptic curve.
We use this example to show that imposing constant intersection numbers gives more information than imposing
self-duality.
Starting from example $4$ we consider Feynman integrals.
Example $4$ illustrates the importance of the symmetrised integrands.
Example $5$ gives the first example, where the entries of the intersection matrix $\overline{\tilde{C}}$ 
are non-trivial Laurent polynomials in $\eps$.
This is an example with Calabi-Yau manifolds.
In example $6$ we study Feynman integrals related to curves of higher genus $g \ge 2$.

\subsection{Example 1}

We start with an example with $\NV=1$, $\NB=3$ and $\NF=2$.
We consider the twist
\bq
 U
 & = & 
 P_0^{-3 \eps} P_1^\eps P_2^\eps P_3^\eps,
\eq
with
\bq
\label{Pi_def}
 P_0 \; =\; z_0,
 \;\;\;\;\;\;
 P_1 \; = \; z_1 - \frac{x_1}{x_0} z_0,
 \;\;\;\;\;\;
 P_2 \; = \; z_1 - \frac{x_2}{x_0} z_0,
 \;\;\;\;\;\;
 P_3 \; = \; z_1 - \frac{x_3}{x_0} z_0.
\eq
We have two master integrands, which can be taken as
\bq
 \check{\differentialform}_1
 & = &
 \differentialform_{1100}\left[1\right]
 \; = \; 
 -3 U \frac{1}{P_0 P_1} \eta,
 \nonumber \\
 \check{\differentialform}_2
 & = &
 \differentialform_{1010}\left[1\right]
 \; = \; 
 -3 U \frac{1}{P_0 P_2} \eta.
\eq
The differential equation for this basis is in $\eps$-factorised form:
\bq
 d_B \left(\begin{array}{c} \check{\differentialform}_1 \\ \check{\differentialform}_2 \\ \end{array} \right)
 & = &
 \eps \check{A} \left(\begin{array}{c} \check{\differentialform}_1 \\ \check{\differentialform}_2 \\ \end{array} \right),
\eq
with
\begin{align}
\label{letters_example_1}
 \check{A}_{11} & = d_B \ln\frac{\left(x_1-x_2\right)\left(x_1-x_3\right)^2}{x_0^3},
 &
 \check{A}_{12} & = d_B \ln\frac{\left(x_1-x_3\right)}{\left(x_1-x_2\right)},
 \\
 \check{A}_{21} & = d_B \ln\frac{\left(x_2-x_3\right)}{\left(x_1-x_2\right)}
 &
 \check{A}_{22} & = d_B \ln\frac{\left(x_1-x_2\right)\left(x_2-x_3\right)^2}{x_0^3}.
\end{align}
The basis for the dual twisted cohomology group is given by
\bq
 \check{\differentialform}_1^\vee
 & = &
 \differentialform_{1100}^\vee\left[1\right]
 \; = \; 
 -3 U^\vee \frac{1}{P_0 P_1} \eta,
 \nonumber \\
 \check{\differentialform}_2^\vee
 & = &
 \differentialform_{1010}^\vee\left[1\right]
 \; = \; 
 -3 U^\vee \frac{1}{P_0 P_2} \eta,
\eq
where
\bq
 U^\vee & = & 
 P_0^{3 \eps} P_1^{-\eps} P_2^{-\eps} P_3^{-\eps}.
\eq
The basis for the dual twisted cohomology group satisfies the differential equation
\bq
\label{dual_differential_equation_example1}
 d_B \left( \check{\differentialform}_1^\vee, \check{\differentialform}_2^\vee \right) & = & - \eps \left( \check{\differentialform}_1^\vee, \check{\differentialform}_2^\vee \right) \check{A}^T.
\eq
The intersection matrix is given by
\bq
 \check{C} & = &
 \frac{1}{\eps} \left( \begin{array}{rr}
 6 & -3 \\
 -3 & 6 \\
 \end{array} \right).
\eq
Although we may express all four entries $\check{A}_{ij}$ as linear combinations of four independent one-forms
\bq
 d_B\ln\left(x_0\right),
 \;\;\;
 d_B\ln\left(x_1-x_2\right),
 \;\;\;
 d_B\ln\left(x_1-x_3\right),
 \;\;\;
 d_B\ln\left(x_2-x_3\right),
\eq
the actual linear combinations, which appear in the differential equation are not independent.
Self-duality gives us the relation
\bq
 \left( \check{A}_{11} - \check{A}_{22} \right) - 2 \left( \check{A}_{12} - \check{A}_{21} \right) & = & 0,
\eq
which is easily verified with the explicit expressions given in eq.~(\ref{letters_example_1}).

\subsection{Example 2}

We keep the divisors as above, but consider now the twist
\bq
 U
 & = & 
 P_0^{-\frac{1}{2}-3 \eps} P_1^{-\frac{1}{2}+\eps} P_2^\eps P_3^\eps.
\eq
Requiring an $\eps$-factorised form will introduce two square roots
\bq
 r_2 \; = \; \sqrt{\frac{x_2-x_1}{x_0}},
 & &
 r_3\; = \; \sqrt{\frac{x_3-x_1}{x_0}}.
\eq
As a basis of master integrands we take
\bq
 \check{\differentialform}_1
 & = &
 r_2 \differentialform_{0010}\left[1\right]
 \; = \; 
 r_2
 U \frac{1}{P_2} \eta,
 \nonumber \\
 \check{\differentialform}_2
 & = &
 r_3 \differentialform_{0001}\left[1\right]
 \; = \; 
 r_3
 U \frac{1}{P_3} \eta.
\eq
The differential equation for this basis is in $\eps$-factorised form:
\bq
 d_B \left(\begin{array}{c} \check{\differentialform}_1 \\ \check{\differentialform}_2 \\ \end{array} \right)
 & = &
 \eps \check{A} \left(\begin{array}{c} \check{\differentialform}_1 \\ \check{\differentialform}_2 \\ \end{array} \right),
\eq
with
\begin{align}
\label{letters_example_2}
 \check{A}_{11} & = d_B \ln\frac{\left(x_1-x_2\right)^2\left(x_2-x_3\right)}{x_0^3},
 &
 \check{A}_{12} & = d_B \ln\frac{1+\frac{r_2}{r_3}}{1-\frac{r_2}{r_3}},
 \\
 \check{A}_{21} & = d_B \ln\frac{1+\frac{r_3}{r_2}}{1-\frac{r_3}{r_2}}
 &
 \check{A}_{22} & = d_B \ln\frac{\left(x_1-x_3\right)^2\left(x_2-x_3\right)}{x_0^3}.
\end{align}
The basis for the dual twisted cohomology group is given by
\bq
 \check{\differentialform}_1^\vee
 & = &
 r_2 \differentialform_{0010}^\vee\left[1\right]
 \; = \; 
 r_2 
 U^\vee \frac{1}{P_2} \eta,
 \nonumber \\
 \check{\differentialform}_2^\vee
 & = &
 r_3 \differentialform_{0001}^\vee\left[1\right]
 \; = \; 
 r_3 
 U^\vee \frac{1}{P_3} \eta,
\eq
where
\bq
 U^\vee & = & 
 P_0^{-\frac{1}{2}+3 \eps} P_1^{-\frac{1}{2}-\eps} P_2^{-\eps} P_3^{-\eps}.
\eq
The basis for the dual twisted cohomology group satisfies a differential equation of the form given in eq.~(\ref{dual_differential_equation_example1}).
The intersection matrix is given by
\bq
 \check{C} & = &
 \frac{1}{\eps} \left( \begin{array}{rr}
 1 & 0 \\
 0 & 1 \\
 \end{array} \right).
\eq
Self-duality gives us again one relation between the entries of the matrix $\check{A}$, which reads
\bq
 \check{A}_{12} - \check{A}_{21} & = & 0.
\eq
Again, this is easily verified from the explicit expressions in eq.~(\ref{letters_example_2}):
\bq
 \check{A}_{12} - \check{A}_{21} 
 & = &
 d_B \ln\frac{\left(1+\frac{r_2}{r_3}\right)\left(1-\frac{r_3}{r_2}\right)}{\left(1-\frac{r_2}{r_3}\right)\left(1+\frac{r_3}{r_2}\right)}
 \; = \;
 d_B \ln\left(-1\right)
 \; = \; 0.
\eq

\subsection{Example 3}

We keep the divisors as above, but consider now the twist
\bq
 U
 & = & 
 P_0^{-\frac{1}{2}-3 \eps} P_1^{-\frac{1}{2}+\eps} P_2^{-\frac{1}{2}+\eps} P_3^{-\frac{1}{2}+\eps}.
\eq
This is now a genus one example. An $\eps$-factorised differential equation for a closely related system was obtained in~\cite{Broedel:2018rwm} and reconsidered in~\cite{Duhr:2024uid} (with two of the ratios $x_i/x_0$ in eq.~(\ref{Pi_def}) ``gauge fixed''). We go through the two steps of the algorithm of refs.~\cite{e-collaboration:2025frv,Bree:2025tug}.
The first step is rational, and does not involve any transcendental functions.
The first step yields an intermediate basis
\bq
 \tilde{\differentialform}_1
 & = &
 \differentialform_{0000}\left[1\right]
 \; = \; 
 U \eta,
 \nonumber \\
 \tilde{\differentialform}_2
 & = &
 \differentialform_{1000}\left[z_1\right]
 \; = \; 
 -\frac{\left(1+6\eps\right)}{2 \eps}
 U \frac{z_1}{P_0} \eta.
\eq
The differential equation for this basis is of the form
\bq
 d_B \left(\begin{array}{c} \tilde{\differentialform}_1 \\ \tilde{\differentialform}_2 \\ \end{array} \right)
 & = &
 \tilde{A} \left(\begin{array}{c} \tilde{\differentialform}_1 \\ \tilde{\differentialform}_2 \\ \end{array} \right),
\eq
with $\tilde{A}$ in Laurent-polynomial form\footnote{In this paper we use the convention that the superscript in brackets refers to the power in the $\eps$-expansion}:
\bq
 \tilde{A}
 & = &
 \left( \begin{array}{rr}
 \tilde{A}^{(0)}_{11} + \eps \tilde{A}^{(1)}_{11} & \eps \tilde{A}^{(1)}_{12} \\
 \frac{1}{\eps} \tilde{A}^{(-1)}_{21} + \tilde{A}^{(0)}_{21} + \eps \tilde{A}^{(1)}_{21} & \tilde{A}^{(0)}_{22} + \eps \tilde{A}^{(1)}_{22} \\
 \end{array} \right).
\eq
The individual entries are given with $x_{ij}=x_i-x_j$ by
\bq
 \tilde{A}^{(1)}_{11}
 & = &
 - \frac{3dx_0}{x_0}
 + \frac{\left(x_1-2x_2-2x_3\right) dx_1}{x_{12} x_{13}}
 + \frac{\left(x_2-2x_1-2x_3\right) dx_2}{x_{21} x_{23}}
 + \frac{\left(x_3-2x_1-2x_2\right) dx_3}{x_{31} x_{32}},
 \nonumber \\
 \tilde{A}^{(1)}_{12}
 & = &
 - \frac{x_0 dx_1}{x_{12} x_{13}}
 - \frac{x_0 dx_2}{x_{21} x_{23}}
 - \frac{x_0 dx_3}{x_{31} x_{32}},
 \nonumber \\
 \tilde{A}^{(1)}_{21}
 & = &
 \frac{3\left(x_1x_2+x_2x_3+x_3x_1\right)}{x_0}
 \left(
 \frac{dx_1}{x_{12} x_{13}}
 + \frac{dx_2}{x_{21} x_{23}}
 + \frac{dx_3}{x_{31} x_{32}}
 \right),
 \nonumber \\
 \tilde{A}^{(1)}_{22}
 & = &
 - \frac{3dx_0}{x_0}
 + \frac{3 x_1 dx_1}{x_{12} x_{13}}
 + \frac{3 x_2 dx_2}{x_{21} x_{23}}
 + \frac{3 x_3 dx_3}{x_{31} x_{32}},
 \nonumber \\
 \tilde{A}^{(0)}_{11}
 & = &
 - \frac{1}{6} \tilde{A}^{(1)}_{22},
 \nonumber \\
 \tilde{A}^{(0)}_{21}
 & = &
   \frac{\left[2x_1\left(x_2+x_3\right)-x_2x_3\right] dx_1}{x_0 x_{12} x_{13}}
 + \frac{\left[2x_2\left(x_1+x_3\right)-x_1x_3\right] dx_2}{x_0 x_{21} x_{23}}
 + \frac{\left[2x_3\left(x_1+x_2\right)-x_1x_2\right] dx_3}{x_0 x_{31} x_{32}},
 \nonumber \\
 \tilde{A}^{(0)}_{22}
 & = &
 \frac{1}{6} \tilde{A}^{(1)}_{22},
 \\
 \tilde{A}^{(-1)}_{21}
 & = &
   \frac{\left[x_1\left(x_2+x_3\right)-x_2x_3\right] dx_1}{4 x_0 x_{12} x_{13}}
 + \frac{\left[x_2\left(x_1+x_3\right)-x_1x_3\right] dx_2}{4 x_0 x_{21} x_{23}}
 + \frac{\left[x_3\left(x_1+x_2\right)-x_1x_2\right] dx_3}{4 x_0 x_{31} x_{32}}.
 \nonumber
\eq
In step $2$ of the algorithm we remove the unwanted terms through a rotation involving transcendental functions:
\bq
 \left(\begin{array}{c} \check{\differentialform}_1 \\ \check{\differentialform}_2 \\ \end{array} \right)
 & = & R_2^{-1}
 \left(\begin{array}{c} \tilde{\differentialform}_1 \\ \tilde{\differentialform}_2 \\ \end{array} \right),
\eq
where $R_2$ is of the form
\bq
 R_2
 & = &
 \left( \begin{array}{rr}
 R^{(0)}_{11} & 0 \\
 \frac{1}{\eps} R^{(-1)}_{21} & R^{(0)}_{22} \\
 \end{array} \right)
 \left( \begin{array}{rr}
 1 & 0 \\
 R^{(0)}_{21} & 1 \\
 \end{array} \right).
\eq
This introduces four ($\eps$-independent)
auxiliary functions $R^{(0)}_{11}$, $R^{(0)}_{22}$, $R^{(-1)}_{21}$ and $R^{(0)}_{21}$, which satisfy the differential equations
\bq
\label{example_3_diff_eq_auxiliary_fcts}
 d_B R^{(0)}_{11} & = & \tilde{A}^{(0)}_{11} R^{(0)}_{11} + \tilde{A}^{(1)}_{12} R^{(-1)}_{21},
 \nonumber \\
 d_B R^{(-1)}_{21} & = & \tilde{A}^{(-1)}_{21} R^{(0)}_{11} + \tilde{A}^{(0)}_{22} R^{(-1)}_{21},
 \nonumber \\
 d_B R^{(0)}_{22} & = & \tilde{A}^{(0)}_{22} R^{(0)}_{22} - R^{(-1)}_{21} \left( R^{(0)}_{11} \right)^{-1} \tilde{A}^{(1)}_{12} R^{(0)}_{22},
 \nonumber \\
 d_B R^{(0)}_{21} & = & \left( R^{(0)}_{22} \right)^{-1} \left( \tilde{A}^{(0)}_{21} R^{(0)}_{11} + \tilde{A}^{(1)}_{22} R^{(-1)}_{21}   
                   - R^{(-1)}_{21} \left( R^{(0)}_{11} \right)^{-1} \tilde{A}^{(1)}_{11} R^{(0)}_{11} \right).
\eq
Explicitly, the basis $(\check{\differentialform}_1,\check{\differentialform}_2)^T$ is given by
\bq
 \check{\differentialform}_1 & = & \frac{\tilde{\differentialform}_1}{R^{(0)}_{11}},
 \nonumber \\
 \check{\differentialform}_2 & = & \frac{\tilde{\differentialform}_2}{R^{(0)}_{22}}
 - \left( \frac{1}{\eps} \frac{R^{(-1)}_{21}}{R^{(0)}_{22}} + R^{(0)}_{21}\right) \frac{\tilde{\differentialform}_1}{R^{(0)}_{11}}.
\eq
The differential equation for this basis is in $\eps$-factorised form.
We now investigate how to eliminate some of the auxiliary functions.
We do this in three ways: first without invoking constant intersection numbers and self-duality,
secondly by imposing constant intersection numbers, and thirdly by imposing self-duality.
\\
\\
{\bf Option 1}: We first discuss the elimination of two auxiliary functions without
any reference to constant intersection numbers or self-duality.
It is easy to check that
\bq
\label{example_3_vanishing_derivatives}
 d_B \left( R^{(0)}_{11} R^{(0)}_{22} \right) & = & 0,
 \nonumber \\
 d_B \left( R^{(0)}_{21} + \frac{\left(x_1+x_2+x_3\right)}{x_0} \frac{R^{(0)}_{11}}{R^{(0)}_{22}} \right) & = & 0,
\eq
hence we may choose the integration constants for $R^{(0)}_{22}$ and $R^{(0)}_{21}$ such that
\bq
\label{example_3_elimination}
 R^{(0)}_{22} & = & \frac{1}{R^{(0)}_{11}},
 \nonumber \\
 R^{(0)}_{21} & = & - \frac{\left(x_1+x_2+x_3\right)}{x_0} \frac{R^{(0)}_{11}}{R^{(0)}_{22}}.
\eq
This eliminates $R^{(0)}_{22}$ and $R^{(0)}_{21}$, in favour of $R^{(0)}_{11}$, and we are left with two auxiliary functions $R^{(0)}_{11}$ and $R^{(-1)}_{21}$, as expected.
$R^{(0)}_{11}$ is a period of the elliptic curve and we may trade $R^{(-1)}_{21}$ for one of its derivatives. The $\eps$-factorised differential equation is independent of $R^{(-1)}_{21}$ and therefore thanks to eq.~(\ref{example_3_elimination}) we conclude that it depends solely on the period, as noted in~\cite{Duhr:2024uid}. It is further easy to check that with the choice of eq.~(\ref{example_3_elimination})
the matrix $\check{A}$ of the $\eps$-factorised differential equation 
satisfies the strong version of self-duality
\bq
 \check{A} \; = \; \persymmetricmatrix \check{A}^T \persymmetricmatrix^{-1}
 & \mbox{where} & \persymmetricmatrix \; = \; \left(\begin{array}{cc} 0 & 1 \\ 1 & 0 \\ \end{array} \right)
\eq
and that the intersection matrix is given by
\bq
 \check{C} & = &
 \frac{1}{\eps} \left( \begin{array}{cc}
 0 & 1 \\
 1 & 0 \\
 \end{array} \right).
\eq
While it is easy to check that eq.~(\ref{example_3_vanishing_derivatives}) is true,
it is not easy to guess which combination to check.
\\
\\
{\bf Option 2}:
We now exploit imposing constant intersection numbers, as is also done in~\cite{Duhr:2024uid}. We will see that this will give us exactly 
the arguments appearing on the left-hand side of eq.~(\ref{example_3_vanishing_derivatives}).
We consider the dual twisted cohomology group and the intersection matrix.
The basis for the dual twisted cohomology group is given by
\bq
 \check{\differentialform}_1^\vee & = & \frac{\tilde{\differentialform}_1^\vee}{R^{(0)}_{11}},
 \nonumber \\
 \check{\differentialform}_2^\vee & = & \frac{\tilde{\differentialform}_2^\vee}{R^{(0)}_{22}}
 - \left( - \frac{1}{\eps} \frac{R^{(-1)}_{21}}{R^{(0)}_{22}} + R^{(0)}_{21}\right) \frac{\tilde{\differentialform}_1^\vee}{R^{(0)}_{11}},
\eq
where
\bq
 \tilde{\differentialform}_1^\vee
 & = &
 \differentialform_{0000}^\vee\left[1\right]
 \; = \; 
 U^\vee \eta,
 \nonumber \\
 \tilde{\differentialform}_2^\vee
 & = &
 \differentialform_{1000}^\vee\left[z_1\right]
 \; = \; 
 \frac{\left(1-6\eps\right)}{2 \eps}
 U^\vee \frac{z_1}{P_0} \eta
\eq
and
\bq
 U^\vee
 & = & 
 P_0^{-\frac{1}{2}+3 \eps} P_1^{-\frac{1}{2}-\eps} P_2^{-\frac{1}{2}-\eps} P_3^{-\frac{1}{2}-\eps}.
\eq
We obtain for the intersection matrix
\bq
 \check{C} & = &
 \frac{1}{\eps} \left( \begin{array}{cc}
 0 & \frac{1}{R^{(0)}_{11} R^{(0)}_{22}} \\
 \frac{1}{R^{(0)}_{11} R^{(0)}_{22}}  & -\frac{2}{R^{(0)}_{11} R^{(0)}_{22}} \left( R^{(0)}_{21} + \frac{\left(x_1+x_2+x_3\right)}{x_0} \frac{R^{(0)}_{11}}{R^{(0)}_{22}} \right) \\
 \end{array} \right).
\eq
If $\check{C}$ is supposed to be constant, there must be constants $\overline{\check{C}}_{12}$ and $\overline{\check{C}}_{22}$ such that
\bq
 R^{(0)}_{11} R^{(0)}_{22} & = & \overline{\check{C}}_{12}^{-1},
 \nonumber \\
 R^{(0)}_{21} + \frac{\left(x_1+x_2+x_3\right)}{x_0} \frac{R^{(0)}_{11}}{R^{(0)}_{22}} & = & - \frac{1}{2} \frac{\overline{\check{C}}_{22}}{\overline{\check{C}}_{12}},
\eq
and eq.~(\ref{example_3_vanishing_derivatives}) follows directly.
\\
\\
{\bf Option 3}:
We now exploit the option of imposing self-duality. The weak version of self-duality states that there is a 
constant symmetric matrix
\bq
 \overline{\check{C}}
 & = & 
 \left( \begin{array}{cc}
  \overline{\check{C}}_{11} & \overline{\check{C}}_{12} \\
  \overline{\check{C}}_{12} & \overline{\check{C}}_{22} \\
 \end{array} \right),
\eq
such that
\bq
 \check{A} \overline{\check{C}} - \overline{\check{C}}^T \check{A}^T & = & 0.
\eq
This gives for this example one equation, which corresponds to eq.~(\ref{relation_Nf_eq_2}).
By series expansion of the entries of $\check{A}$ we can establish that
\bq
 \overline{\check{C}}_{11} \; = \; 0,
 \;\;\;
 \overline{\check{C}}_{12} \; = \; 1,
 \;\;\;
 \overline{\check{C}}_{22} \; = \; 0
\eq
is an allowed solution.
In this case, eq.~(\ref{relation_Nf_eq_2}) reduces to
\bq
 R^{(0)}_{21} + \frac{\left(x_1+x_2+x_3\right)}{x_0} \frac{R^{(0)}_{11}}{R^{(0)}_{22}} & = & 0.
\eq
We note that imposing just self-duality misses the relation
\bq
\label{example_3_eq_1}
 R^{(0)}_{11} R^{(0)}_{22} & = & 1.
\eq
This is not surprising: In the one-variable Calabi-Yau case (e.g. the equal mass $l$-loop banana integral)
self-duality is the statement that the $Y$-invariants satisfy $Y_j=Y_{\NV+1-j}$ \cite{2013arXiv1304.5434B,Pogel:2022vat}.
In addition we have $Y_1=1$, which corresponds to eq.~(\ref{example_3_eq_1}).
We conclude that imposing constant intersection numbers gives more information.

\subsection{Example 4}

\begin{figure}
\begin{center}
\includegraphics[scale=1.0]{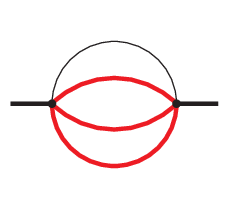}
\end{center}
\caption{
A three-loop contribution to the electron self-energy.
Massive propagators are indicated by red lines.
}
\label{fig:example_4}
\end{figure}

We consider example 5.5 from ref.~\cite{Bree:2025tug}.
The example corresponds to a specific three-loop contribution to the electron self-energy~\cite{Duhr:2024bzt},
the three-loop banana graph with one zero mass.
The graph is shown in fig.~\ref{fig:example_4}.
The twist function is given by
\bq 
 U\left(z_0,z_1,z_2\right)
 & = &
 P_0^{4\eps}
 P_1^\eps 
 P_2^{-2\eps}
 P_3^{-\frac{1}{2}}
 P_4^{-\frac{1}{2}-\eps}
 P_5^{-\frac{1}{2}-\eps},
\eq
with
\begin{align}
 P_0 & = z_0, & P_3 & = z_1, \nonumber \\
 P_1 & = z_2, & P_4 & = z_1 + 4 x z_0, \nonumber \\
 P_2 &= z_2+z_0, & P_5 & = \left(z_2-z_1\right)^2 +2 x z_0 \left( z_1+z_2\right) + x^2 z_0^2,
\end{align}
where in the notation we use here $x=x_1/x_0$.
The dimension of the twisted cohomology group is $\dim \Hgen^{2} = 6$, whereas the dimension on the Feynman integral side
is $\dim V^2=3$; the difference in dimensions is due to two symmetry relations and one master integrand in a super-sector.

We first decompose the six-dimensional basis for $\Hgen^2$ into basis elements corresponding to symmetry relations, 
basis elements corresponding to the sector of interest and basis elements corresponding to super-sectors.
We use the basis
\begin{align}
 \mbox{symmetry} & : 
 &
 \tilde{\differentialform}_1 & = 
 6x \differentialform_{000010}\left[1\right]
 + \differentialform_{001000}\left[1\right]
 + \differentialform_{100000}\left[1\right],
 \nonumber \\
 & &
 \tilde{\differentialform}_2 & = 
 4\left(1+3x\right) \differentialform_{001000}\left[1\right]
 + 2 \differentialform_{100000}\left[1\right]
 + 3 \differentialform_{101000}\left[z_1\right],
 \nonumber \\
 \mbox{sector} & :
 & 
 \tilde{\differentialform}_3 & =
 \differentialform_{001000}\left[1\right], 
 \nonumber \\
 & &
 \tilde{\differentialform}_4 & =
 \differentialform_{100000}\left[1\right],
 \nonumber \\
 & &
 \tilde{\differentialform}_5 & =
 \differentialform_{001010}\left[z_0\right]
 - \frac{1}{48x\left(x-1\right)} \left( 16 \tilde{\differentialform}_1 + \tilde{\differentialform}_2 \right),
 \nonumber \\
 \mbox{super-sector} & : &
 \tilde{\differentialform}_6 & =
 \differentialform_{010000}\left[1\right].
\end{align}
In this basis the corresponding matrix $\tilde{A}^{\mathrm{twist}}$ of the differential equation $d_B \tilde{\differentialform} = \tilde{A}^{\mathrm{twist}} \tilde{\differentialform}$ is of the form
\bq
\label{six_times_six_matrix}
 \tilde{A}^{\mathrm{twist}}
 & = &
 \left( \begin{array}{cccccc}
 * & * & 0 & 0 & 0 & 0 \\
 * & * & 0 & 0 & 0 & 0 \\
 0 & 0 & 0 & 0 & * & 0 \\
 0 & 0 & * & * & 0 & 0 \\
 0 & 0 & * & * & * & 0 \\
 0 & 0 & * & 0 & 0 & * \\
 \end{array} \right),
\eq
where the symbol $*$ indicates a non-zero entry.
In this basis, the derivatives of the integrands of the sector
$\tilde{\differentialform}_3, \dots, \tilde{\differentialform}_5$ do not
involve the integrands $\tilde{\differentialform}_1, \tilde{\differentialform}_2$, which correspond
to the symmetry relations.
In particular, the entries $\tilde{A}_{ij}^{\mathrm{twist}}$ with $i \in \{3,4,5\}$ and $j \in \{1,2\}$ are zero.

At the level of Feynman integrals we obtain
after step $1$ of the algorithm of refs.~\cite{e-collaboration:2025frv,Bree:2025tug} the basis
\bq
 \left( \begin{array}{c}
 J_1 \\
 J_2 \\
 J_3 \\
 \end{array} \right)
 & = &
 \left( \begin{array}{c}
  \left\langle \differentialform_{001000}\left[1\right] | {\mathcal C}_{\mathrm{Feynman}} \right\rangle \\
  \left\langle \differentialform_{100000}\left[1\right] | {\mathcal C}_{\mathrm{Feynman}} \right\rangle \\
  \left\langle \differentialform_{001010}\left[z_0\right] | {\mathcal C}_{\mathrm{Feynman}} \right\rangle \\
 \end{array} \right).
\eq
The differential equation reads $d_B J = \tilde{A} J$, where
\bq
 \tilde{A}
 & = & 
 \left( \begin{array}{ccc}
 0 & 0 & 6 \eps \\
 - \frac{4\eps}{x} & - \frac{4\eps}{x} & 0 \\
 \tilde{A}_{31} & - \frac{\eps}{x^2\left(x-1\right)\left(9x-1\right)} & \tilde{A}_{33} \\
 \end{array} \right),
\eq
with
\bq
 \tilde{A}_{31}
 & = & 
 - \frac{3x-1}{2x\left(x-1\right)\left(9x-1\right)\eps} - \frac{21x-5}{2x\left(x-1\right)\left(9x-1\right)} - \frac{\left(18x^2-3x+1\right)\eps}{x^2\left(x-1\right)\left(9x-1\right)},
 \nonumber \\
 \tilde{A}_{33}
 & = & 
 - \frac{27x^2-20x+1}{x\left(x-1\right)\left(9x-1\right)} - \frac{\left(63x^2-30x-1\right)\eps}{x\left(x-1\right)\left(9x-1\right)}.
\eq
The matrix $\tilde{A}$ is the $(3 \times 3)$-submatrix 
formed by the rows and columns $\{3,4,5\}$ of the $(6 \times 6)$-matrix $\tilde{A}^{\mathrm{twist}}$ in eq.~(\ref{six_times_six_matrix}).
From the discussion of section~\ref{sect:compatibility} it follows that 
\bq
 \left\langle \differentialform_{001000}^{\mathrm{sym}}\left[1\right] \right|
 & = & 
 \left\langle \differentialform_{001000}\left[1\right] \right|,
 \nonumber \\
 \left\langle \differentialform_{100000}^{\mathrm{sym}}\left[1\right] \right|
 & = &
 \left\langle \differentialform_{100000}\left[1\right] \right|,
 \nonumber \\
 \left\langle \differentialform_{001010}^{\mathrm{sym}}\left[z_0\right] \right|
 & = &
 \left\langle \differentialform_{001010}\left[z_0\right] \right|
 - \frac{1}{48x\left(x-1\right)} \left( 16 \left\langle \tilde{\differentialform}_1 \right| + \left\langle \tilde{\differentialform}_2 \right| \right).
\eq
We see that symmetrisation acts trivially on
$\left\langle \differentialform_{001000}\left[1\right] \right|$ and $\left\langle \differentialform_{100000}\left[1\right] \right|$,
but non-trivially on $\left\langle \differentialform_{001010}\left[z_0\right] \right|$.

Integrating the differential equation for the rescaled intersection matrix eq.~(\ref{diff_eq_overlineCtilde}) we obtain
\bq
\label{intersection_Ctilde_example_4}
 \overline{\tilde{C}}
 & = &
 f\left(\eps\right)
 \left( \begin{array}{ccc}
 0 & 0 & \frac{1}{2x\left(x-1\right)\left(9x-1\right)} \\
 0 & 2 & 0 \\
 \frac{1}{2x\left(x-1\right)\left(9x-1\right)} & 0 & \frac{1+30x-63x^2}{12 x^2\left(x-1\right)^2\left(9x-1\right)^2} \\
 \end{array} \right).
\eq
This gives $\tilde{C}$ up to an unknown prefactor $f(\eps)$.
The prefactor is of no relevance in the following and we may set this prefactor to one (i.e. $f(\eps)=1$).
This ensures that $\det \overline{\tilde{C}}$ is independent of $\eps$.

In step two of the algorithm of refs.~\cite{e-collaboration:2025frv,Bree:2025tug}
we rotate the system to an $\eps$-form with the rotation matrix
\bq
 R_2 & = & R_2^{(-1)} R_2^{(0)}.
\eq
The ansatz for $R_2^{(-1)}$ and $R_2^{(0)}$ introduces four auxiliary functions
\begin{align}
 R_2^{(-1)}
 & = 
 \left( \begin{array}{cc|c}
  R^{(0)}_{11} & 0 & 0  \\
  0 & 1 & 0 \\
 \hline
  \frac{1}{\eps} R^{(-1)}_{31} & 0 & R^{(0)}_{33} \\
 \end{array} \right),
 &
 R_2^{(0)}
 & = 
 \left( \begin{array}{cc|c}
  1 & 0 & 0 \\
  0 & 1 & 0 \\
 \hline
  R^{(0)}_{31} & 0 & 1 \\
 \end{array} \right).
\end{align}
We then compute
\bq
 \overline{\check{C}} & = & R_2^{-1} \overline{\tilde{C}} \left(R_2^{\vee \; T}\right)^{-1}.
\eq
This yields
\bq
 \overline{\check{C}} & = &
 \left( \begin{array}{ccc}
 0 & 0 & \frac{1}{2x\left(x-1\right)\left(9x-1\right) R^{(0)}_{11} R^{(0)}_{33}} \\
 0 & 2 & 0 \\
 \frac{1}{2x\left(x-1\right)\left(9x-1\right) R^{(0)}_{11} R^{(0)}_{33}} & 0 & \frac{1+30x-63x^2}{12 x^2\left(x-1\right)^2\left(9x-1\right)^2 (R^{(0)}_{33})^2 } - \frac{R^{(0)}_{31}}{x\left(x-1\right)\left(9x-1\right)R^{(0)}_{11} R^{(0)}_{33}}\\
 \end{array} \right).
\eq
We then verify
\bq
 d_B \left( x\left(x-1\right)\left(9x-1\right) R^{(0)}_{11} R^{(0)}_{33} \right) & = & 0,
 \nonumber \\
 d_B \left( R^{(0)}_{31} - \frac{\left(1+30x-63x^2\right)R^{(0)}_{11}}{12 x\left(x-1\right)\left(9x-1\right) R^{(0)}_{33}} \right) & = & 0.
\eq
This allows us to eliminate $R^{(0)}_{33}$ and $R^{(0)}_{31}$ by setting
\bq
 R^{(0)}_{33} & = & \frac{1}{2x\left(x-1\right)\left(9x-1\right) R^{(0)}_{11}},
 \nonumber \\
 R^{(0)}_{31} & = & \frac{1}{6} \left(1+30x-63x^2\right) \left(R^{(0)}_{11}\right)^2.
\eq
For the rescaled intersection matrix $\overline{\check{C}}$ we obtain
\bq
 \overline{\check{C}} & = &
 \left( \begin{array}{ccc}
 0 & 0 & 1 \\
 0 & 2 & 0 \\
 1 & 0 & 0 \\
 \end{array} \right).
\eq

\subsection{Example 5: The Calabi-Yau case}
\label{sect:example_5}

We consider the $l$-loop equal-mass banana integral.
This is the prime example of a Feynman integral related 
to a Calabi-Yau $(l-1)$-fold.
It is also one of the first examples where it has been observed that some of the auxiliary functions
can be eliminated thanks to the strong version of self-duality \cite{Pogel:2022yat,Pogel:2022ken,Pogel:2022vat}. The elimination of auxiliary functions for deformed Calabi-Yau operators~\cite{Duhr:2025lbz}, using the constancy of the intersection matrix, was also considered in Ref.~\cite{Duhr:2025xyy}, in context of the approach of~\cite{Gorges:2023zgv,Duhr:2025lbz}.

We follow the notation of ref.~\cite{Pogel:2022vat}
and we set $x=m^2/(-p^2)$.
On the maximal cut we have $l$ master integrals.
We may start on the maximal cut with the derivative basis
\bq
\label{def_derivative_basis}
 J & = &
 \left(
 I_{1 \dots 1 1},
 \;\;\;
 \frac{1}{\eps} \frac{d}{dx}I_{1 \dots 1 1},
 \;\;\;
 \dots,
 \;\;\;
 \frac{1}{\eps^{l-1}} \frac{d^{l-1}}{dx^{l-1}}I_{1 \dots 1 1}
 \right)^T.
\eq
The differential equation for this basis is compatible with the $\Fcomb$-filtration. 
We have one master integral in every non-trivial part of the filtration:
\bq
 \emptyset = \Fcomb^{l} \Hgen^{l-1} \subseteq \Fcomb^{l-1} \Hgen^{l-1} \subseteq \dots \subseteq \Fcomb^1 \Hgen^{l-1} \subseteq \Fcomb^0 \Hgen^{l-1} = \Hgen^{l-1}.
\eq
We look at the rescaled intersection matrix $\overline{\tilde{C}}$ for the basis $J$,
which we obtain by integrating the differential equation in eq.~(\ref{diff_eq_overlineCtilde}).
The choice $f(\eps)=1$ for the prefactor is sufficient for the $\eps$-independence of $\det \overline{\tilde{C}}$.
We find that all entries are Laurent polynomials in $\eps$, with the highest degree in $\eps$ being $\eps^0$
and the lowest degree in $\eps$ given by
\bq
 \mathrm{ldegree}
 & = &
 \left\{ \begin{array}{ll}
  -l+1, & \mbox{$l$ odd}, \\
  -l+2, & \mbox{$l$ even}. \\
 \end{array}
 \right.
\eq
In order to understand the structure of the intersection matrix $\overline{\tilde{C}}$, we list the lowest degrees of
the various entries for $l \in \{1,\dots,5\}$.
\bq
 l=1 & : &
 \mathrm{ldegree}\left(\overline{\tilde{C}},\eps\right)
 \; = \;
 \left( \begin{array}{r}
 0 \\
 \end{array} \right),
 \nonumber \\
 l=2 & : &
 \mathrm{ldegree}\left(\overline{\tilde{C}},\eps\right)
 \; = \;
 \left( \begin{array}{rr}
 - & 0 \\
 0 & 0 \\
 \end{array} \right),
 \nonumber \\
 l=3 & : &
 \mathrm{ldegree}\left(\overline{\tilde{C}},\eps\right)
 \; = \;
 \left( \begin{array}{rrr}
 - & - & 0 \\
 - & 0 & -1 \\
 0 & -1 & -2 \\
 \end{array} \right),
 \nonumber \\
 l=4 & : &
 \mathrm{ldegree}\left(\overline{\tilde{C}},\eps\right)
 \; = \;
 \left( \begin{array}{rrrr}
 - & - & - & 0 \\
 - & - & 0 & -1 \\
 - & 0 & 0 & -2 \\
 0 & -1 & -2 & -2 \\
 \end{array} \right),
 \nonumber \\
 l=5 & : &
 \mathrm{ldegree}\left(\overline{\tilde{C}},\eps\right)
 \; = \;
 \left( \begin{array}{rrrrr}
 - & - & - & - & 0 \\
 - & - & - & 0 & -1 \\
 - & - & 0 & -1 & -2 \\
 - & 0 & -1 & -2 & -3 \\
 0 & -1 & -2 & -3 & -4 \\
 \end{array} \right).
\eq
A ``$-$'' indicates that the corresponding intersection number is zero.
The pattern is easily understood as follows: The lowest degree in $\eps$ is bounded by $(-l+1)$.
This bound is reached for $l$ being odd.
For $l$ being even, the bound is not saturated. The reason is given by the symmetry properties discussed 
in section~\ref{sect:symmetry_intersection_matrix}.
For $l$ even, terms proportional to $\eps^{-l+1}$ must be anti-symmetric and can therefore not appear on the diagonal.

\begin{figure}
\begin{center}
\includegraphics[scale=1.0]{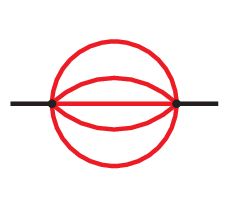}
\end{center}
\caption{
The four-loop equal mass banana integral. Red lines are massive propagators.
}
\label{fig:example_5}
\end{figure}

We now discuss the case $l=4$ in more detail. This is the first case where a non-trivial $Y$-invariant appears.
The graph is shown in fig.~\ref{fig:example_5}.
In step 2 of the algorithm of refs.~\cite{e-collaboration:2025frv,Bree:2025tug} we go from the filtration-compatible basis $J$ to the $\eps$-factorised basis $K$ by a rotation 
\bq
 R_2 & = & R_2^{(-3)} R_2^{(-2)} R_2^{(-1)} R_2^{(0)},
\eq
where
\begin{align}
\label{R2_four_loop_banana}
 R_2^{(-3)}
 & = 
 \left( \begin{array}{cccc}
  {\color{red} R^{(0)}_{11}} & 0 & 0 & 0 \\
  \frac{1}{\eps} {\color{red} R^{(-1)}_{21}} & {\color{red} R^{(0)}_{22}} & 0 & 0 \\
  \frac{1}{\eps^2} {\color{red} R^{(-2)}_{31}} & \frac{1}{\eps} {\color{red} R^{(-1)}_{32}} & {\color{blue} R^{(0)}_{33}} & 0 \\
  \frac{1}{\eps^3} {\color{red} R^{(-3)}_{41}} & \frac{1}{\eps^2} {\color{blue} R^{(-2)}_{42}} & \frac{1}{\eps} {\color{blue} R^{(-1)}_{43}} & {\color{blue} R^{(0)}_{44}} \\
 \end{array} \right),
 &
 R_2^{(-2)}
 & = 
 \left( \begin{array}{cccc}
  1 & 0 & 0 & 0 \\
  {\color{red} R^{(0)}_{21}} & 1 & 0 & 0 \\
  \frac{1}{\eps} {\color{red} R^{(-1)}_{31}} & {\color{blue} R^{(0)}_{32}} & 1 & 0 \\
  \frac{1}{\eps^2} {\color{blue} R^{(-2)}_{41}} & \frac{1}{\eps} {\color{blue} R^{(-1)}_{42}} & {\color{blue} R^{(0)}_{43}} & 1 \\
 \end{array} \right),
 \nonumber \\
 R_2^{(-1)}
 & = 
 \left( \begin{array}{cccc}
  1 & 0 & 0 & 0 \\
  0 & 1 & 0 & 0 \\
  {\color{red} R^{(0)}_{31}} & 0 & 1 & 0 \\
  \frac{1}{\eps} {\color{red} R^{(-1)}_{41}} & {\color{blue} R^{(0)}_{42}} & 0 & 1 \\
 \end{array} \right),
 &
 R_2^{(0)}
 & = 
 \left( \begin{array}{cccc}
  1 & 0 & 0 & 0 \\
  0 & 1 & 0 & 0 \\
  0 & 0 & 1 & 0 \\
  {\color{blue} R^{(0)}_{41}} & 0 & 0 & 1 \\
 \end{array} \right).
\end{align}
This introduces $20$ auxiliary functions.
The requirement of a constant intersection matrix allows us to express $10$ of those auxiliary functions
(shown in blue in eq.~(\ref{R2_four_loop_banana}))
as algebraic expressions of the remaining $10$ auxiliary functions
(shown in red in eq.~(\ref{R2_four_loop_banana})).
An example is given by the relation
\bq
 R^{(0)}_{44}
 & = &
 \frac{1}{x\left(x+1\right)\left(9x+1\right)\left(25x+1\right)R^{(0)}_{11}}.
\eq
The remaining $10$ auxiliary functions are determined by a system of first-order differential equations.
As we can always convert a system of first-order differential equations to higher-order differential equations
for single functions, we may express the remaining $10$ auxiliary functions in terms 
of four auxiliary functions and derivatives thereof.
These four auxiliary functions can be chosen as $R^{(0)}_{11}$, which satisfies a homogeneous fourth-order 
differential equation, $R^{(0)}_{22}$, which satisfies a homogeneous second-order 
differential equation and $R^{(0)}_{21}$, $R^{(0)}_{31}$, which satisfy inhomogeneous second-order differential
equations.
The function $R^{(0)}_{11}$ is a period of the Calabi-Yau three-fold, the function $R^{(0)}_{22}$
is related to the Jacobian of the mirror map.

We elaborate on the requirement to choose appropriate boundary values for the auxiliary functions.
It is easily shown that
\bq
 d_B \overline{\check{C}}^{(-2)}_{34} & = & 0.
\eq
We may use this equation to eliminate $R^{(-2)}_{42}$. The general solution is
\bq
 R^{(-2)}_{42}
 & = &
 - N^{(-2)}_{34} x \left(x+1\right)\left(9x+1\right)\left(25x+1\right) R^{(0)}_{22} R^{(0)}_{33} R^{(0)}_{44}
 + \frac{R^{(-1)}_{32}R^{(-1)}_{21}}{R^{(0)}_{11}}
 - \frac{R^{(0)}_{22}R^{(-2)}_{31}}{R^{(0)}_{11}}
 \nonumber \\
 & &
 - \frac{1+70x+777x^2+900x^3}{x\left(x+1\right)\left(9x+1\right)\left(25x+1\right)} R^{(-1)}_{32}
 - \frac{1+28x+285x^2+450x^3}{x^2\left(x+1\right)\left(9x+1\right)\left(25x+1\right)} R^{(0)}_{22},
\eq
where $N^{(-2)}_{34}$ is a boundary constant.
We then have
\bq
 \overline{\check{C}}^{(-2)}_{34} & = & N^{(-2)}_{34}.
\eq
It is clear that $N^{(-2)}_{34}=0$ is the appropriate choice, as we want to eliminate all terms proportional to $\eps^{-2}$ in the intersection matrix $\overline{\check{C}}$.
However, let us investigate what happens if we choose $N^{(-2)}_{34} \neq 0$.
We then have
\bq
 d_B \overline{\check{C}}^{(-1)}_{24} & = & N^{(-2)}_{34} \frac{R^{(0)}_{33}}{R^{(0)}_{22}} dx.
\eq
We cannot have $R^{(0)}_{33}=0$, as $\det R_2$ must be non-zero. Hence, we see that the intersection matrix $\overline{\check{C}}$
will not be constant for $N^{(-2)}_{34} \neq 0$.

\subsection{Example 6: Curves of higher genus}

\begin{figure}
\begin{center}
\includegraphics[scale=1.0]{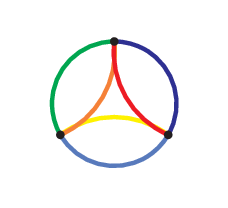}
\hspace*{10mm}
\includegraphics[scale=1.0]{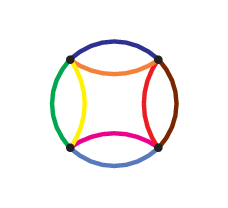}
\hspace*{10mm}
\includegraphics[scale=1.0]{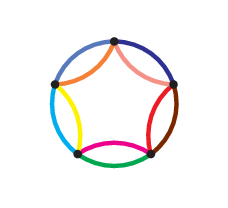}
\end{center}
\caption{
The four-loop, the five-loop and the six-loop necklace diagrams. All propagators have pairwise distinct non-zero masses.
}
\label{fig:example_6}
\end{figure}

The prime examples of Feynman integrals related to curves of higher genus are the necklace integrals with unequal masses.
The necklace diagrams with four, five and six loops are shown in fig.~\ref{fig:example_6}.
The $l$-loop necklace integral is related to a curve of genus $g=l-2$.

For the four-loop necklace diagram we have six master integrals in the top sector.
The Hodge-like diagram is given by
\begin{center}
\begin{axopicture}(280,140)(0,0)
\Text(110,60)[c]{$2$}
\Text(150,60)[c]{$2$}
\Text(130,80)[c]{$2$}
\DashLine(30,70)(260,70){6}
\DashLine(30,90)(260,90){6}
\Text(253,60)[c]{$W_1$}
\Text(253,80)[c]{$W_2$}
\Line(260,70)(260,64)
\Line(260,90)(260,84)
\DashLine(120,10)(220,110){3}
\DashLine(80,10)(180,110){3}
\Text(110,20)[c]{$\Fgeom^0$}
\Text(70,20)[c]{$\Fgeom^1$}
\Line(120,10)(117,10)
\Line(80,10)(77,10)
\end{axopicture}
\end{center}
Two master integrals are related to the two holomorphic one-forms of the genus two curve and span $\Hgen^{(1,0)}$, 
while two other masters are related to residues at points and span $\Hgen^{(1,1)}$.
Finally, the remaining two master integrals are related to two non-holomorphic one-forms of the genus two curve and span $\Hgen^{(0,1)}$.
We look at the rescaled intersection matrix $\overline{\tilde{C}}$ for the basis $J=(J_1,J_2,J_3,J_4,J_5,J_6)^T$,
where
\bq
 J_1, J_2 & \in & \Hgen^{(1,0)},
 \nonumber \\
 J_3, J_4 & \in & \Hgen^{(1,1)},
 \nonumber \\
 J_5, J_6 & \in & \Hgen^{(0,1)}.
\eq
We find that all entries are Laurent polynomials in $\eps$, with the highest degree in $\eps$ being $\eps^0$
and the lowest degree in $\eps$ given by
\bq
\label{ldegree_necklace_4}
 \mathrm{ldegree}\left(\overline{\tilde{C}},\eps\right)
 \; = \;
 \left( \begin{array}{rr|rr|rr}
 - & - & - & - & 0 & 0 \\
 - & - & - & - & 0 & 0 \\
 \hline
 - & - & 0 & - & 0 & 0 \\
 - & - & - & 0 & 0 & 0 \\
 \hline
 0 & 0 & 0 & 0 & 0 & -1 \\
 0 & 0 & 0 & 0 & -1 & 0 \\
 \end{array} \right).
\eq
In eq.~(\ref{ldegree_necklace_4}) we indicated in addition the decomposition with respect to $\Hgen^{(p,q)}$ by lines.
For curves of genus $g \ge 2$ we find that the bound of $\mathrm{ldegree}=-1$ is reached, as we can have an anti-symmetric
sub-matrix in the lower right corner.
In the special case of $g=1$ (discussed in the previous example in section~\ref{sect:example_5}) the bound is not reached, as there cannot be
an anti-symmetric $(1 \times 1)$-sub-matrix.

\section{Conclusions}
\label{sect:conclusions}

In this paper we investigated intersection matrices associated to geometric-ordered bases of Feynman integrals.
The intersection numbers are defined for the integrands of Feynman integrals.
In order to have a meaningful interpretation at the level of Feynman integrals, independent of the integral representation we work with, we need to consider symmetrised integrands.
The intersection matrix for the symmetrised integrands is most efficiently computed from a differential equation.
With these preparations we considered the intersection matrices on the maximal cut of a
filtration-compatible basis and an $\eps$-factorised basis, as obtained after step $1$ and step $2$
of the algorithm of refs.~\cite{e-collaboration:2025frv,Bree:2025tug}, respectively.
We find that these intersection matrices are simpler than expected:
For a filtration-compatible basis the entries of the intersection matrix are Laurent polynomials 
in the dimensional regularisation parameter $\varepsilon$,
for an $\varepsilon$-factorised basis the entries of the intersection matrix are integers,
if an overall prefactor given as a power of $\varepsilon$ is factored out
and if the boundary values for the auxiliary functions are chosen appropriately.
Turning the argument around we may require an $\eps$-factorised differential equation and 
constant intersection numbers.
There are indications that a constant intersection matrix plays an important role in the context of $\eps$-factorised differential equations \cite{Duhr:2024xsy}.
We showed that the algorithm of refs.~\cite{e-collaboration:2025frv,Bree:2025tug}
allows us to construct a basis, which has constant intersection numbers on the maximal cut.
We expect this to be true beyond the maximal cut, but a verification of this property would require the machinery of relative
twisted cohomology.

On the practical side, our findings can be used to eliminate some of the auxiliary functions introduced in step $2$ of the
algorithm of refs.~\cite{e-collaboration:2025frv,Bree:2025tug}.
We presented a systematic algorithm, 
which eliminates algebraically redundant auxiliary functions on the maximal cut
and which minimises the required calculations.
The elimination of redundant auxiliary functions will be important in concrete precision calculations.

\subsection*{Acknowledgements}

We thank Matthias Wilhelm for spotting a few typos in the manuscript. 
FG wishes to thank Claude Duhr, Sara Maggio, Cathrin Semper and Sven F. Stawinski for discussions. The work of FG is supported by the European Union (ERC Consolidator Grant LoCoMotive $101043686$). Views and 
opinions expressed are however those of the author(s) only and do not necessarily 
reflect those of the European Union or the European Research Council. Neither the 
European Union nor the granting authority can be held responsible for them.
X.W. is supported by the University Development Fund of The Chinese University of Hong Kong, Shenzhen, under the Grant No. UDF01003912.
X.W. is also supported in part by the National Natural Science Foundation of China with Grant No. 12535006.


\begin{appendix}

\section{The map to compact support}
\label{sect:map_iota}

In this appendix we discuss in more detail the map to compact support 
appearing in the definition of the intersection numbers
\bq
 C_{ij}
 \;= \;
 \left\langle \differentialform_i \right. \left| \differentialform_j^\vee \right\rangle
 & = &
 \frac{C_iC_j^\vee}{\left(2\pi i\right)^n}
 \int
  \iota_\omega\left(\myhat{\Phi}_i \eta\right)
  \wedge
  \iota_{-\omega}\left(\frac{\myhat{\Phi}_j^\vee \eta}{\Divisor_{\mathrm{odd}}}\right).
\eq
For simplicity, we focus on the case $\NV=1$ and we follow closely the discussion in ref.~\cite{Mizera:2019gea,Mizera:2019ose}.
In this case, the map $\iota_\omega$ is explicitly given by
\bq
 \iota_\omega\left(\myhat{\Phi}_i \eta\right)
 & = & 
 \myhat{\Phi}_i \eta - \nabla_\omega \sum\limits_{z' \in D} \Theta\left(\delta_1^2-\left|z-z'\right|^2\right) \nabla_\omega^{-1} \left( \myhat{\Phi}_i \eta \right),
\eq
where $\delta_1$ is an arbitrary infinitesimal positive constant.
In a similar way, we may map the second form to compact support:
\bq
 \iota_{-\omega}\left(\frac{\myhat{\Phi}_j^\vee \eta}{\Divisor_{\mathrm{odd}}}\right)
 & = &
 \frac{\myhat{\Phi}_j^\vee \eta}{\Divisor_{\mathrm{odd}}} - \nabla_{-\omega} \sum\limits_{z' \in D} \Theta\left(\delta_2^2-\left|z-z'\right|^2\right) \nabla_{-\omega}^{-1} \left( \frac{\myhat{\Phi}_j^\vee \eta}{\Divisor_{\mathrm{odd}}} \right),
\eq
where $\delta_2$ is another arbitrary infinitesimal positive constant.
We may choose $\delta_1 \neq \delta_2$. Such a choice will avoid complications due to the square of a delta-distribution.
Without loss of generality we assume $\delta_1 > \delta_2$.
In this case, $\iota_\omega\left(\myhat{\Phi}_i \eta\right)$ has no support on $\left|z-z'\right|^2=\delta_2^2$, while
$\iota_{-\omega}(\Divisor_{\mathrm{odd}}^{-1} \myhat{\Phi}_j^\vee \eta)$ is identical to $(\Divisor_{\mathrm{odd}}^{-1}\myhat{\Phi}_j^\vee \eta)$
on $\left|z-z'\right|^2=\delta_1^2$. Hence, the formula for the intersection number reduces to
\bq
 C_{ij}
 \;= \;
 \left\langle \differentialform_i \right. \left| \differentialform_j^\vee \right\rangle
 & = &
 \frac{C_iC_j^\vee}{\left(2\pi i\right)^n}
 \int
  \iota_\omega\left(\myhat{\Phi}_i \eta\right)
  \wedge
  \left(\frac{\myhat{\Phi}_j^\vee \eta}{\Divisor_{\mathrm{odd}}}\right).
\eq
The choice $\delta_1 < \delta_2$ gives the alternative formula
\bq
 C_{ij}
 \;= \;
 \left\langle \differentialform_i \right. \left| \differentialform_j^\vee \right\rangle
 & = &
 \frac{C_iC_j^\vee}{\left(2\pi i\right)^n}
 \int
  \left(\myhat{\Phi}_i \eta\right)
  \wedge
  \iota_{-\omega}\left(\frac{\myhat{\Phi}_j^\vee \eta}{\Divisor_{\mathrm{odd}}}\right).
\eq

\section{Invariance of intersection numbers}
\label{sect:invariance}

In this appendix we provide more information on eq.~(\ref{rescaling_equation}):
\bq
 \frac{1}{\left(2\pi i\right)^n}
 \int \iota_\omega\left(\myhat{\Phi}_i \eta\right) \wedge \iota_{-\omega}\left(\frac{\myhat{\Phi}_j^\vee \eta}{\Divisor_{\mathrm{odd}}}\right)
 & = &
 \frac{1}{\left(2\pi i\right)^n}
 \int \iota_{\omega+\kappa}\left(\frac{\myhat{\Phi}_i \eta}{T}\right) \wedge \iota_{-\omega-\kappa}\left(\frac{T \myhat{\Phi}_j^\vee \eta}{\Divisor_{\mathrm{odd}}}\right).
\eq
We first note that
in order to compute intersection numbers, we split the differential forms 
$\differentialform_i$ into 
a (multi-valued) twist function $U$ and a (rational single-valued) differential form
$C_i \myhat{\Phi}_i \eta$.
In general, this operation is only defined in the ambient space ${\mathbb C}^{\NV+1}$.
Only if the degree of homogeneity of $U$ with respect to the variables $z$ is zero (e.g $d_U=0$), then this splitting
is well-defined in projective space ${\mathbb C}{\mathbb P}^{\NV}$.

We may use eq.~(\ref{rescaling_equation}) for a specific $T^{\mathrm{spec}}$ such that the degree of homogeneity of $U^{\mathrm{hom}}=U T^{\mathrm{spec}}$ with respect to the variables $z$ is zero.
In this case, the twist function $U^{\mathrm{hom}}$ as well as the rational differential forms
$(T^{\mathrm{spec}})^{-1} \myhat{\Phi}_i \eta$ and $T^{\mathrm{spec}} \Divisor_{\mathrm{odd}}^{-1} \myhat{\Phi}_j^\vee \eta$ are all defined on 
${\mathbb C}{\mathbb P}^{\NV}$.
In this way we may define intersection numbers for a splitting defined in the ambient space ${\mathbb C}^{\NV+1}$ in terms of quantities all defined 
in projective space ${\mathbb C}{\mathbb P}^{\NV}$.

Let $U^{\mathrm{hom}}$ now be a twist function homogeneous of degree zero with respect to the variables $z$.
We set $\omega^{\mathrm{hom}} = d_F \ln U^{\mathrm{hom}}$ and we denote
\bq
 \myhat{\Phi}_{\mathrm{left}} \eta \; = \; \frac{\myhat{\Phi}_i \eta}{T^{\mathrm{spec}}},
 & &
 \myhat{\Phi}_{\mathrm{right}} \eta \; = \; \frac{T^{\mathrm{spec}} \myhat{\Phi}_j^\vee \eta}{\Divisor_{\mathrm{odd}}}.
\eq
Both $\myhat{\Phi}_{\mathrm{left}} \eta$ and $\myhat{\Phi}_{\mathrm{right}} \eta$ are
homogeneous of degree zero with respect to the variables $z$.
We may now consider transformations $T^{\mathrm{hom}}$ of the form as in eq.~(\ref{def_T_shift}) and in addition 
homogeneous of degree zero with respect to the variables $z$.
We set $\kappa^{\mathrm{hom}} = d_F \ln T^{\mathrm{hom}}$.
The weak form of eq.~(\ref{rescaling_equation}) reads
\bq
\label{rescaling_equation_weak_version}
\lefteqn{
 \frac{1}{\left(2\pi i\right)^n}
 \int \iota_{\omega^{\mathrm{hom}}}\left(\myhat{\Phi}_{\mathrm{left}} \eta\right) \wedge \iota_{-\omega^{\mathrm{hom}}}\left(\myhat{\Phi}_{\mathrm{right}} \eta\right)
 = } & &
 \nonumber \\
 & &
 \frac{1}{\left(2\pi i\right)^n}
 \int \iota_{\omega^{\mathrm{hom}}+\kappa^{\mathrm{hom}}}\left(\frac{\myhat{\Phi}_{\mathrm{left}} \eta}{T^{\mathrm{hom}}}\right) \wedge \iota_{-\omega^{\mathrm{hom}}-\kappa^{\mathrm{hom}}}\left(T^{\mathrm{hom}} \myhat{\Phi}_{\mathrm{right}} \eta\right).
\eq
In this formula all quantities are defined in projective space ${\mathbb C}{\mathbb P}^{\NV}$
and this formula has appeared in the literature before \cite{Chestnov:2022xsy,Matsubara-Heo:2023hmf}.

We finally note that the intersection numbers do not depend how we achieve a twist function with $d_U=0$.
To this aim let $T^{\mathrm{spec}}_1$ and $T^{\mathrm{spec}}_2$ be two transformations, which lead to
$U^{\mathrm{hom}}_1=U T^{\mathrm{spec}}_1$ and $U^{\mathrm{hom}}_2=U T^{\mathrm{spec}}_2$, respectively, with
\bq
 d_{U^{\mathrm{hom}}_1} \; = \; d_{U^{\mathrm{hom}}_2} \; = \; 0.
\eq
We also define
\bq
 \myhat{\Phi}_{\mathrm{left},1} \eta \; = \; \frac{\myhat{\Phi}_i \eta}{T^{\mathrm{spec}}_1},
 & &
 \myhat{\Phi}_{\mathrm{right},1} \eta \; = \; \frac{T^{\mathrm{spec}}_1 \myhat{\Phi}_j^\vee \eta}{\Divisor_{\mathrm{odd}}},
 \nonumber \\
 \myhat{\Phi}_{\mathrm{left},2} \eta \; = \; \frac{\myhat{\Phi}_i \eta}{T^{\mathrm{spec}}_2},
 & &
 \myhat{\Phi}_{\mathrm{right},2} \eta \; = \; \frac{T^{\mathrm{spec}}_2 \myhat{\Phi}_j^\vee \eta}{\Divisor_{\mathrm{odd}}},
\eq
We have
\bq
 U^{\mathrm{hom}}_2 \; = \; T^{\mathrm{hom}} U^{\mathrm{hom}}_1
 & \mbox{with} &
 T^{\mathrm{hom}} \;= \; T^{\mathrm{spec}}_2 \left( T^{\mathrm{spec}}_1 \right)^{-1}.
\eq
$T^{\mathrm{hom}}$ is homogeneous of degree zero with respect to the variables $z$ and eq.~(\ref{rescaling_equation_weak_version}) applies.
Hence
\bq
 \frac{1}{\left(2\pi i\right)^n}
 \int \iota_{\omega^{\mathrm{hom}}_1}\left(\myhat{\Phi}_{\mathrm{left},1} \eta\right) \wedge \iota_{-\omega^{\mathrm{hom}}_1}\left(\myhat{\Phi}_{\mathrm{right},1} \eta\right)
 & = &
 \frac{1}{\left(2\pi i\right)^n}
 \int \iota_{\omega^{\mathrm{hom}}_2}\left(\myhat{\Phi}_{\mathrm{left},2} \eta\right) \wedge \iota_{-\omega^{\mathrm{hom}}_2}\left(\myhat{\Phi}_{\mathrm{right},2} \eta\right).
 \nonumber \\
\eq

\end{appendix}

\bibliography{/home/stefanw/notes/biblio}
\bibliographystyle{h-physrev5}

\end{document}